\documentclass[12pt,english,floatfix,superscriptaddress,aps,prd,preprint,showkeys]{revtex4}
\usepackage{amsmath}
\usepackage{amssymb}
\usepackage{amsbsy}
\usepackage{amsfonts}
\usepackage{amsopn}
\usepackage{amstext}
\usepackage{graphicx}
\usepackage{amssymb}
\usepackage{amsfonts}
\usepackage{amsmath}
\usepackage{graphicx}
\usepackage[english]{babel}
\usepackage{color}
\usepackage{slashed}
\usepackage{esint}
\usepackage[dvips]{epsfig}
\usepackage[dvips]{graphicx}
\usepackage{float}
\usepackage{units}
\usepackage{textcomp}

\usepackage{hyperref}
\usepackage{slashed}
\usepackage{hyperref}
\usepackage{graphicx}
\usepackage{amssymb}
\usepackage{amsmath}
\usepackage{color}

\begin{document}


\title{Thick string-like braneworlds in f(T) gravity}


\author{A. R. P. Moreira}
\affiliation{Universidade Federal do Cear\'a (UFC), Departamento de F\'isica,\\ Campus do Pici, Fortaleza - CE, C.P. 6030, 60455-760 - Brazil.}

\author{J. E. G. Silva}
\email{euclides.silva@ufca.edu.br}
\affiliation{Universidade Federal do Cariri(UFCA), Av. Tenente Raimundo Rocha, \\ Cidade Universit\'{a}ria, Juazeiro do Norte, Cear\'{a}, CEP 63048-080, Brasil}

\author{D.F.S.Veras}
\affiliation{N\'{u}cleo de Tecnologia, Centro Universit\'{a}rio Christus (Unichristus), Campus Dom Lu\'{i}s - Av. Dom Lu\'{i}s, 911, Aldeota, CEP 60160-196, Fortaleza, Cear\'{a}, Brazil.}


\author{C.A.S. Almeida}
\affiliation{Universidade Federal do Cear\'a (UFC), Departamento de F\'isica,\\ Campus do Pici, Fortaleza - CE, C.P. 6030, 60455-760 - Brazil.}

\begin{abstract}
We propose a codimension two warped braneworld model within the teleparallel $f(T)$ gravity. By assuming a global vortex as the source, we found a $l=0$ vortex solution that yields to a thick string-like brane. Asymptotically, the bulk geometry converges to an $AdS_6$ spacetime whose cosmological constant is produced by the torsion parameters. Furthermore, the torsion induces an $AdS-dS$ transition on the exterior region. Inside the brane core, the torsion produces an internal structure even for a single complex scalar field. As the torsion parameters vary, the brane undergoes a phase transition leading to the formation of ring-like structures. The bulk-brane Planck mass ration is modified by the torsion. The analysis of the stress energy condition reveals a splitting brane process satisfying the weak and strong energy conditions for some values of the parameters. In addition, we investigate the behaviour of the gravitational perturbations in this scenario. It turns out that the gravitational spectrum has a linear behaviour for small masses and is independent of the torsion parameters for large masses. In the bulk, the torsion keeps a gapless non-localizable and stable tower of massive modes. Inside the brane core the torsion produces new barriers and potential wells leading to small amplitude massive modes and a massless mode localized around the ring structures. 
\end{abstract}


\maketitle

\section{Introduction}

The Randall-Sundrum models (RS) opened new possibilities for the geometry of the multidimensional spacetime \cite{rs,rs2}.
By assuming a warped geometry, the propagation of the gravitational \cite{Csaki1}, gauge \cite{Kehagias}, and fermionic fields \cite{CASA} are governed by the bulk curvature. Furthermore, the bulk curvature allows geometric solutions for the gauge hierarchy \cite{rs2}, dark matter \cite{darkmatter} and the cosmological constant problems \cite{cosmologicalconstant}.

In five dimensions, domain-walls branes offer stable solutions whose width and internal structure steam from the self-interaction potential and the coupling to the geometry \cite{domainwall}. In a general relativity based theory, wherein the gravitational sector is governed by the Einstein-Hilbert action, the brane internal structure is driven by the interaction of two scalar fields \cite{baseia2}. On the other hand, by modifying the gravitational interaction, thick brane solutions can also be found in a pure geometric setup. Notably, in a Weyl geometry, where the metric compatibility condition is violated by a scalar field, the Weyl scalar field also produces thick branes with internal structure \cite{weylgeometry1,weylgeometry2}. Other modified gravity theories provides further deformed solutions, as in $f(R)$ \cite{fR1,fR2} and mimetic gravity \cite{mimetic}.

Amidst the modified gravity theories, the teleparallel $f(T)$ theories have attracted much attention recently \cite{ft}. Such models assume that the gravitational interaction is encoded not in the curvature but in the torsion \cite{Aldrovandi}. Instead of the metric, the dynamical variables in teleparallelism are the vielbein, and a special connection, the so-called Weitenzb\"{o}ck connection, ensures the vanishing of the curvature tensor \cite{Linder}. Unlike the $f(R)$ models, teleparallel $f(T)$ theories leads to second-order equations of the motion (eom) \cite{ftpalatini}. In cosmology, $f(T)$ models furnish geometric solutions to the early inflation \cite{Ferraroinflation} and to the late acceleration phase \cite{ftdarkenergy1,ftdarkenergy2}. In spite of the extra degrees of freedom present in the vielbeins \cite{ftdegreeoffreedom}, the linear perturbation upon the flat spacetimes has only two propagating degrees of freedom \cite{ftgw,ftgw2,ftgw3}. The modifications of the $f(T)$ gravity was also explored on black holes \cite{Miao,blackhole1,blackhole2}, binary objects \cite{ftbinary} and violation of local Lorentz symmetry \cite{lorentzviolation}. 

The effects of the torsion have been studied in braneworlds models, both in the Einstein-Cartan \cite{torsion1,torsion2}  and in the teleparallel gravities \cite{Capozziello,KKft1,KKft2}. In a codimension one warped model, a power-law $f(T)$ teleparallel gravity with a single real scalar field leading to a thick domain-wall braneworld with internal structure was found \cite{Yang2012}. This solution shares properties akin to those generated by two interacting scalar fields in a GR based theory. Indeed, the torsion parameters modify the profile of the energy density leading to the splitting into two branes \cite{Yang2012}. The torsion modifications were extended by allowing general coupling to scalar fields \cite{Menezes,ftnoncanonicalscalar,ftborninfeld} and a mimetic theory \cite{ftmimetic}. In a metric formulation, the $f(T)$ gravity induces changes on the stress-energy tensor, thereby modifying the source equation of state \cite{ftenergyconditions}. The torsion also changes the dynamics of bulk fermions and the gravitational perturbations \cite{tensorperturbations}. 

In this work we propose a $f(T)$ teleparallel codimension two braneworld. In six dimensions, axially symmetric solutions of the GR based Einstein equations, known as the string-like branes, exhibit a rich geometric and physical properties \cite{gregory,Gherghetta,Oda,Giovannini:2001hh}.
The internal manifold formed by the two extra dimensions has intrinsic properties, such as the deficit angle, related to the brane features \cite{Liu,apple,Papantonopoulos,burgess,conifold,cigar, regularstring}. Moreover, the two extra dimensions enables the localization of the gauge fields considering only a minimal geometric coupling \cite{gaugestring}.

Assuming a global vortex formed by a single complex scalar field we obtained a smooth thick string-like brane whose width is controlled by the torsion. By varying the torsion parameters, the source undergoes a phase transition revealed by the stress energy components. The profile of the energy density and the pressures show that the torsion tends to split the brane, forming a ring-like structure. In the thin string-like regime, the torsion provides a source for the bulk cosmological constant, thereby leading to a warped compactification. The gravitational perturbations form a gaplass Kaluza-Klein (KK) spectrum whose interaction with the brane depends on the torsion parameters.

The work is organized as follows. In section (\ref{sec1}) we review the main definitions of the teleparallel $f(T)$ theory and build the respective string-like braneworld. Furthermore, the exterior and interior solutions are found, as well as the vortex solution. We examine the stress energy tensor components and the brane tensions. In section (\ref{sec4}) we derive the tensor perturbed equations and explore the gravitational KK modes. Finally, additional comments are discussed in section (\ref{finalremarks}). 

\section{Teleparallel braneworld}
\label{sec1}

In this section we present the main concepts of the teleparallel $f(T)$ gravity and obtain the modified gravitational equations
for the braneworld scenario.

In teleparallel gravity, the dynamic variable is provided by the vielbeins, defined by $g_{MN}=\eta_{ab}h^a\ _M h^b\ _N$, where the capital latin index $M={0,...,D-1}$ are the bulk coordinate indexes and the latin index $a={0,...,D-1}$ is a vielbein index.
In order to allow a distant parallelism, the teleparallel gravity assumes a curvature free connection, known as the Weitzenb\"{o}ck connection, defined by  
\cite{Aldrovandi}
\begin{eqnarray}\label{a.685}
\widetilde{\Gamma}^P\ _{NM}=h_a\ ^P\partial_M h^a\ _N.
\end{eqnarray}
The Weitzenb\"{o}ck connection has a non vanishing torsion, defined as $T^{P}\  _{MN}= \widetilde{\Gamma}^P\ _{NM}-\widetilde{\Gamma}^\P\ _{MN}$. The Weitzenb\"{o}ck and the torsion free connections are related by $\widetilde{\Gamma}^P\ _{NM}= \Gamma^P\ _{NM} + K^P\ _{NM}$
where $K^P\ _{NM}=( T_N\ ^P\ _M +T_M\ ^P\ _N - T^P\ _{NM})/2$ is the contorsion tensor \cite{Aldrovandi}.

By defining the so-called superpotential torsion tensor as $S_{P}\ ^{MN}=( K^{MN}\ _{P}-\delta^N_P T^{QM}\ _Q+\delta^M_P T^{QN}\ _Q)/2$, 
a teleparallel equivalent gravity Lagrangian reads $\mathcal{L}=-\kappa_g hT/4$, 
where $h=\sqrt{g}$, with $g$ the determinant of the metric and $T=T^{P}\  _{MN} T_{P}\ ^{MN}/2 +T^{P}\ _{MN}T^{NM}\ _{P}-2T^{P}\ _{MP}T^{NM}\ _{N}=T_{PMN}S^{PMN}$ is a quadratic torsion invariant \cite{Aldrovandi}. Such Lagrangian is equivalent to the usual Einstein-Hilbert action. Indeed, the Ricci scalar for the Weitzenb\"{o}ck is proportional to $T$ by $R=-T+\nabla_{M}T^{MN}_{N}$ \cite{Aldrovandi}.
 
A modified gravity theory can be accomplished by considering as the gravitational Lagrangian a function of the quadratic torsion invariant $f(T)$ \cite{ft,Aldrovandi}. We assume a six dimensional bulk $f(T)$ teleparallel gravity in the form  
\begin{eqnarray}\label{55.5}
\mathcal{S}=-\frac{1}{4\kappa_g}\int h f(T)d^6x+\int \left(\Lambda +\mathcal{L}_m\right)d^6x,
\end{eqnarray}
where $\kappa_g$ is the gravitational constant and $\mathcal{L}_m$ is the matter Lagrangian.
The modified gravity field equation has the form \cite{Yang2012}
\begin{eqnarray}\label{3.36}
\frac{1}{h}f_T\left(\partial_Q\left(h S_N\ ^{MQ}\right)-h\widetilde{\Gamma}^R\ _{SN}S_R\ ^{MS}\right)-f_{TT}S_N\ ^{MQ}\partial_Q T+\frac{1}{4}\delta_N^Mf=-\kappa_g(\Lambda\delta_N^M+\mathcal{T}_N\ ^M),
\end{eqnarray}
where $f\equiv f(T)$, $f_T\equiv\partial f(T)/\partial T$ e $f_{TT}\equiv\partial^2 f(T)/\partial T^2$. An equivalent metric gravitational field equation is given by \cite{ftgw}
\begin{eqnarray}
\label{equivalentmetric}
R_{MN}-\frac{1}{2}Rg_{MN}&=&\frac{\kappa_g}{ f_T}\mathcal{T}_{MN}+\mathfrak{T}_{MN},
\end{eqnarray}
where the source-like term provided by the torsion has the form
\begin{equation}
\mathfrak{T}_{MN}=\left[[ f_T - f(T)]g_{MN}-f_{TT} S_{MNP}\nabla^{P}T\right]/f_T.
\end{equation}
Therefore, the $f(T)$ torsion effects are equivalent to a GR additional source and a varying gravitational constant. Note that, for a constant torsion scalar $T$, the vacuum spacetime is equivalent to one with a cosmological constant of form
\begin{equation}
\label{cosmologicalconstant}
\Lambda_T = -\left[ T - \frac{ f(T)}{f_T} \right]. 
\end{equation}
 
We seek for a codimension two axisymmetric braneworld. A suitable metric ansatz for the string-like braneworld is given by \cite{Gherghetta,Liu}
\begin{equation}\label{45.a}
ds^2=e^{2A(r)}\eta_{\mu\nu}dx^\mu dx^\nu+dr^2+R^2_0e^{2B(r)}d\theta^2,
\end{equation}
where $0 \leq r \leq r_{max} $, $\theta \in [0; 2\pi)$ and $e^{A(r)}$ is the so-called warp factor. The geometry is smooth at the origin provided that \cite{conifold,cigar}
\begin{eqnarray}
\label{regularityconditions}
e^{A(0)}=1 &,& (e^{A})'(0)=0\nonumber\\
e^{B(0)}=0 &,& (e^{B})'(0)=1.
\end{eqnarray}
Accordingly, adopting the \textit{sechsbeins} in the form $h_a\ ^\mu=diag(e^A, e^A, e^A, e^A, 1, R_0 e^B)$, the torsion scalar is given by $T=-4A'\left(3A'+2B'\right)$, where the prime $(\ '\ )$ denotes differentiation with respect to $r$. 

The correction on the gravitational constant provided by the $f(T)$ in Eq.(\ref{equivalentmetric}) yields to a modification on the relationship between the bulk and the brane Planck masses as
\begin{equation}
\label{planckmasses}
M_{4}^{2}=2\pi R_0 M_{6}^{4} \int_{0}^{\infty}{f_T e^{2A+B}dr}.
\end{equation}

For an axisymmetric source we assume a global vortex whose Lagrangian is given by
\begin{equation}
\mathcal{L}_M=-h\left[\frac{1}{2}\partial^M\Phi\partial_M\Phi^*+V(\Phi)\right],
\end{equation}
where $\Phi=\phi(r)e^{il\theta}$ is a complex scalar field. The corresponding stress energy tensor is
\begin{equation}
\mathcal{T}_{MN}=-\frac{1}{2}\left(\partial_M\Phi\partial_N\Phi^*+\partial_M\Phi^*\partial_N\Phi\right)+\frac{1}{2}g_{MN}g^{PQ}\partial_P\Phi\partial_Q\Phi^*+g_{MN}V,    
\end{equation}
whose components with respect to the metric (\ref{45.a}) are
\begin{eqnarray}\label{q.0}
&& t_0= \frac{1}{2}\left[\phi'^2+R_0^{-2}e^{-2B}(l\phi)^2\right]+V,\nonumber\\
&& t_\theta= -\frac{1}{2}\left[\phi'^2-R_0^{-2}e^{-2B}(l\phi)^2\right]+V,\nonumber\\
&& t_r = \frac{1}{2}\left[\phi'^2-R_0^{-2}e^{-2B}(l\phi)^2\right]+V,
\end{eqnarray}
where, $\mathcal{T}_{\mu}\ ^{\rho}=t_{0} \delta^{\rho}_{\mu}$, $\mathcal{T}_\theta\ ^\theta=t_\theta $ and $ \mathcal{T}_r\ ^r=t_r $.

In this work we consider a power-law modified gravity in the form $f(T)=T+kT^n$, where $k$ and $n$ are two torsion parameters controlling the departure of the usual teleparallel theory \cite{Yang2012}. The corresponding equations of motion are 
\begin{eqnarray}
\label{scalarfieldeom}
\frac{1}{2}\left[{\Phi^*}''+(4A'+B'){\Phi^*}'-R_0^{-2}e^{-2B}l^2\Phi^*\right]=\frac{\partial V}{\partial \Phi}
\end{eqnarray}

\begin{eqnarray}\label{e.1}
\frac{1}{2}\left\{1+4^{n-1}nk\left[-A'\left(3A'+2B'\right)\right]^{n-1}\right\}\left[\left(3A'+B'\right)\left(4A'+B'\right)+3A''+B''\right]\nonumber\\ -4^{n-1}kn\left(n-1\right)\left(3A'+B'\right)\left[-A'\left(3A'+2B'\right)\right]^{n-2}\left[A''B'+A'\left(3A''+B''\right)\right]\nonumber\\
-3A'^2-2A'B'+4^{n-1}k\left[-A'\left(3A'+2B'\right)\right]^n=-\kappa_g\Big\{\Lambda+\frac{1}{2}\left[\phi'^2+R_0^{-2}e^{-2B}(l\phi)^2\right]+V\Big\},
\end{eqnarray}
\begin{eqnarray}\label{e.2}
2\left\{1+4^{n-1}nk\left[-A'\left(3A'+2B'\right)\right]^{n-1}\right\}\left[A'\left(3A'+B'\right)+A''\right]\nonumber\\ -4^{n-1}kn\left(n-1\right)A'\left[-A'\left(3A'+2B'\right)\right]^{n-2}\left[A''B'+A'\left(3A''+B''\right)\right]\nonumber\\
-3A'^2-2A'B'+4^{n-1}k\left[-A'\left(3A'+2B'\right)\right]^n= -\kappa_g\Big\{\Lambda-\frac{1}{2}\left[\phi'^2-R_0^{-2}e^{-2B}(l\phi)^2\right]+V\Big\} ,
\end{eqnarray}
\begin{eqnarray}\label{e.3}
\left(-4\right)^{n-1}k\left(2n-1\right)\left[A'\left(3A'+2B'\right)\right]^{n}&\nonumber\\
+A'\left(3A'+2B'\right)=&-\kappa_g\Big\{\Lambda+\frac{1}{2}\left[\phi'^2-R_0^{-2}e^{-2B}(l\phi)^2\right]+V\Big\}.
\end{eqnarray}
The equations (\ref{scalarfieldeom}), (\ref{e.1}), (\ref{e.2}) and (\ref{e.3}) form a quite intricate system of coupled equations. Thus, we first analyse the solutions exterior to the brane and then, we propose a possible solution for the brane core.  

\subsection{Thin string like regime}

We assume as the exterior brane the region where both the potential and the field derivative $\phi'$ vanish. This solution can also represent the bulk geometry of a thin string-like brane \cite{Gherghetta}.  By restricting our analysis to the $l=0$ field configuration, the vacuum exterior geometry is governed a bulk cosmological constant term. Thus, assuming $\mathcal{T}_N\ ^M=0$  and $A'=B'=-c$, the EOMs (\ref{e.1}), (\ref{e.2}) e (\ref{e.3}) yields to
\begin{eqnarray}\label{e.1.a}
5c^2+\left(-4\right)^{n-1}k\left(2n-1\right)(5c^2)^{n}= -\kappa_g\Lambda.
\end{eqnarray}
The Eq.(\ref{e.1.a}) establishes a relation between the bulk cosmological constant, the torsion parameters and $c$. The corresponding exterior metric takes the form
\begin{eqnarray}\label{a.65}
ds^2=e^{-2cr}\eta_{\mu\nu}dx^\mu dx^\nu + dr^2 + R^{2}_{0} e^{-2cr} d\theta^{2},
\end{eqnarray}
which is the thin string-like brane solution in the Gherghetta-Schaposhnikov (GS) model \cite{Gherghetta}.
However, unlike the usual GR braneworld, the torsion modification enables new possible configurations.

Firstly, let us seek for solutions of Eq.(\ref{e.1.a}) for $\Lambda=0$. If $n=1$ the only solution is $c=0$, which leads to a factorizable Kaluza-Klein model. Yet, for $n=2$ we obtain the solution
\begin{eqnarray}
\label{cosmologicalconstant1}
c=\pm \sqrt{\frac{\kappa_g}{60 k}},
\end{eqnarray}
whereas for $n=3$ we find $c=\pm \sqrt[4]{\frac{1}{200k}}$. Accordingly, the torsion yields to a warped compactified spacetime even in the absence of a bulk cosmological constant. Indeed, from Eq.(\ref{cosmologicalconstant}) and Eq.(\ref{cosmologicalconstant1}) we obtain an effective cosmological constant as
\begin{eqnarray}
\label{cosmologicalconstanttorsion}
\Lambda_T =-\frac{1}{3k}.
\end{eqnarray}
Since $k>0$ in Eq.(\ref{cosmologicalconstant1}), the torsion produces a negative cosmological constant.

Now let us consider a non-zero bulk cosmological constant and analyse how the torsion modifies the Eq.(\ref{cosmologicalconstant1}).
For $n=1$, the Eq.(\ref{e.1.a}) yields to
\begin{eqnarray}
\label{cn1}
c= \pm\sqrt{\frac{\kappa_g\left(-\Lambda\right)}{5\left(1+k\right)}}.
\end{eqnarray}
The Eq.(\ref{cn1}) reveals a striking effect of torsion upon the exterior geometry.
For $\Lambda<0$, the torsion parameter can assume the values in the range $k>-1$, whereas for $k<-1$ the bulk cosmological constant should be positive. This transition from a $AdS_6$ into a $dS_6$ spacetime when $k$ goes from $k>-1$ to $k<-1$ can be 
summarized by the modified effective bulk cosmological constant $\Lambda_{mod}$ in the form
\begin{eqnarray}
\Lambda_{mod}=\frac{\Lambda}{k+1}.
\end{eqnarray}
It is worthwhile to mention that $\Lambda_{mod}>\Lambda$ if $-1<k<0$.
Moreover, the $AdS-dS$ transition is discontinuous at $k=-1$. 

For $n=2$ we obtain four solutions
\begin{eqnarray}
c= \pm\sqrt{\frac{\kappa_g\left(-\Lambda\right)}{5}}\quad or \quad  c= \pm\sqrt{\frac{\kappa_g\left(\Lambda+1\right)}{60k}}.
\end{eqnarray}
The first two solutions represent the usual thin string-like brane with a $AdS_6$ bulk whereas the last two solutions are generated by a positive cosmological constant provided $k>0$. 
Thus, for both $n=2$ and $n=1/2$ the exterior geometry is the same as in the thin GS model. Note that in the range $0< k<1$, the torsion increases the constant $c$ when compared to the thin string solution \cite{Gherghetta}.

In the thin string brane limit the relationship between the Planck masses is modified by the torsion as shown in Eq.(\ref{planckmasses}). For instance, if $n=2$ in the absence of the bulk cosmological constant, the relation reads
\begin{eqnarray}
\label{thinplanckmassrelation}
M^{2}_{4}=-\frac{20\pi\mu_\theta}{3\Lambda_T}\left(1-\frac{20k\Lambda_{T}}{3}\right) M_{6}^{4},
\end{eqnarray}
where $\mu_\theta =2R_0 c M_{6}^{4}$ is the thin string like angular tension \cite{Gherghetta}. For $M_4 \gg M_6$, we set $\Lambda_T\ll \mu_\theta$. It is worthwhile to mention that torsion correction in Eq.(\ref{thinplanckmassrelation}) is bigger than the GS relation by the factor $k\Lambda_T$.

Note that for $A'=B'=-c$, the torsion invariant $T$ is constant of form $T=-20 c^2$. Likewise the GR braneworld vacuum has constant Ricci scalar, the teleparallel vacuum solution has constant torsion invariant. The particular choice of the $f(T)$ modifies the value and the sign of this torsion invariant. 

A noteworthy comment on the Eq.(\ref{e.1.a}) is the appearance of complex solutions for half-integer values of $n$. For 
$\Lambda=0$, the only real solution allowed is the trivial one $c=0$. On the other hand, for $\Lambda\neq 0$ the complex solution 
$c=c_1 + i c_2$ leads to the real warp factor $Re(e^{2A})=e^{-c_1 r}\cos c_2 r$. A similar behaviour was found in stationary solutions called standing wave braneworld \cite{standingwave}. Henceforward, we restrict our analysis to integers $n$. 

\subsection{{Thick string-like brane}}
\label{sec3}

Once we studied the effects of the torsion on the exterior geometry, we seek for smooth thick string-like solutions whose warp factor has the form \cite{Yang2012}
\begin{eqnarray}
\label{coreA}
e^{2A(r)}=\cosh^{-2p}(\lambda r),
\end{eqnarray}
where the parameters $p$ and $\lambda$ determine, respectively, the amplitude and the width of the source. For the angular warp factor we assume the ansatz 
\begin{eqnarray}
\label{coreB}
B(r)=\ln[\cosh^{-b}(\lambda r)]+\ln[\tanh(\rho r)],
\end{eqnarray}
where the second term guarantees the regularity condition at the origin (\ref{regularityconditions}). The metric components (\ref{coreA}) and (\ref{coreB}) bears a resemblance to the GR based thick string-like branes \cite{conifold,cigar,regularstring}.  
The corresponding infinitesimal element reads
\begin{eqnarray}
\label{coremetric}
ds^2 = e^{2A}[\eta_{\mu\nu}dx^\mu dx^\nu + dr^2 + R_{0}^2 \tanh^2 (\rho r)d\theta^2],
\end{eqnarray} 
which is the warped product between a $3-$brane and a cigar internal manifold \cite{cigar}. For $r\geq Max(\lambda,\rho)$, the geometry (\ref{coremetric}) recovers the exterior solution (\ref{a.65}). 

\begin{figure}
\begin{center}
\begin{tabular}{ccc}
\includegraphics[height=5cm]{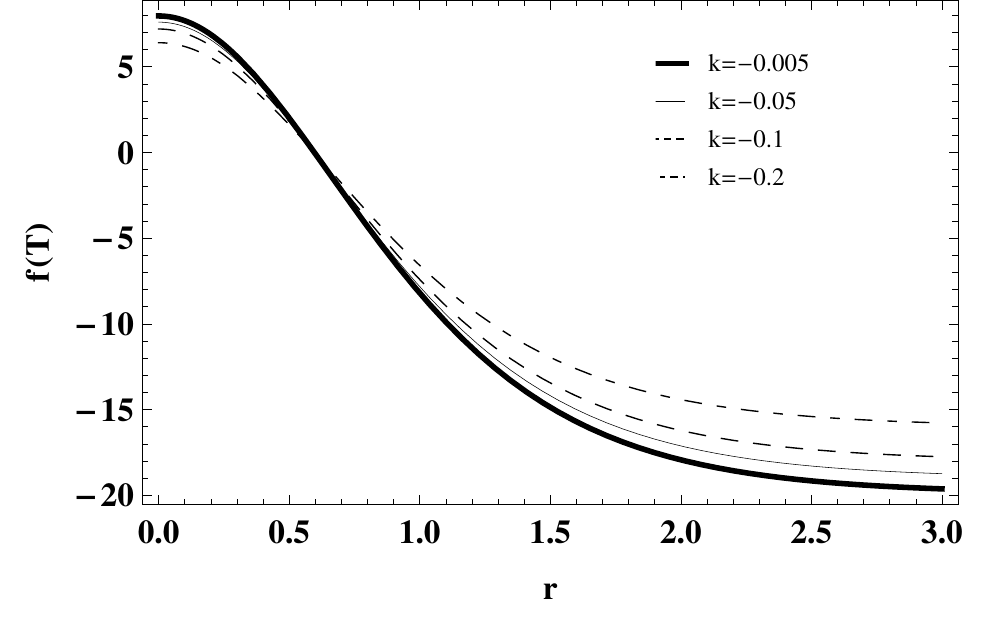}
\includegraphics[height=5cm]{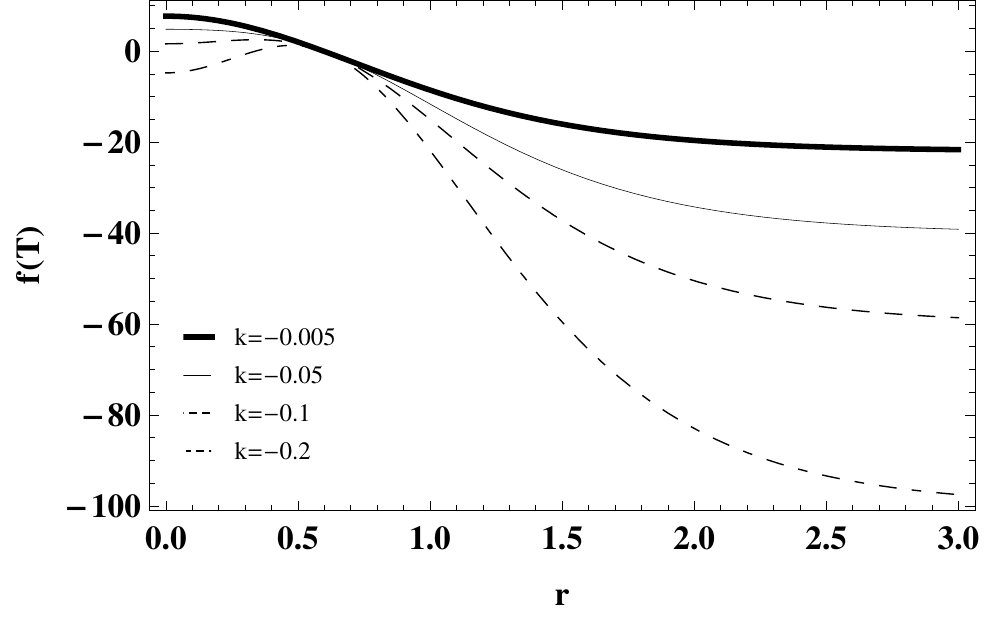}\\
(a) \hspace{8 cm}(b)
\end{tabular}
\end{center}
\caption{$f(T)$  for $\rho=\lambda=p=1$. (a) $n=1$.
(b)  $n=2$. 
\label{figf(T)1}}
\end{figure}

Let us first study the geometry features of this solution. In Fig. (\ref{figf(T)1}), we plotted the $f(T)$ function for this thick solution for $n=1$ and $n=2$. For $n=1$, the torsion invariant is concentrated around the origin and decays to a constant value asymptotically. The $n=2$ configuration exhibits a maximum value displayed from the origin and a higher asymptotic value. 
Thus, the torsion parameters yield to geometric modifications inside the brane core.

\subsubsection{Vortex solution}

Now let us turn out attention to the complex scalar field source. Following the approach carried out in Ref. \cite{Yang2012}, the modified Einstein equations yields to
\begin{eqnarray}\label{q.4}
-4\left(A'^2-7A'B'-B'^2+5A''-B''\right)
+4^{n-2}[-A'(3A'+2B')]^{n-2}kn\Bigg\{2(n-1)B'^2A''\nonumber\\ -24A'^4+2A'B'\left[B'^2-(2n+3)A''+nB'\right]-15A'^2A''+5B'A'^3+17A'^2B'^2+3A'^2B''
\nonumber\\ 
-10(n-1)A'^2(3A''+B'')\Bigg\}=-2\phi'^2+R_0^{-2}e^{-2B}(l\phi)^2,
\end{eqnarray}
where we set the gravitational constant $\kappa_g=1$ for simplicity.   The Eq.(\ref{q.4}) allows us to obtain the scalar field for a given geometric solution \cite{Yang2012}. For $l=0$, $\lambda=1$, $\rho=1$ e $p=1$, the scalar field satisfies

\begin{eqnarray}\label{q.5}
\frac{7\ \mathrm{sech}^2(r)}{16[9-5\ \cosh^2(2r)]}\Bigg\{748-720\ \cosh(2r)+100 \cosh(4r)&\nonumber\\
+  4^nkn[7\ \mathrm{sech}^2(r)-5]^n\left[13+8n \cosh(2r)+(8n-13)\cosh(4r)\right]\Bigg\}&=2\phi'^2.
\end{eqnarray}

\begin{figure}
\begin{center}
\begin{tabular}{ccccccccc}
\includegraphics[height=5cm]{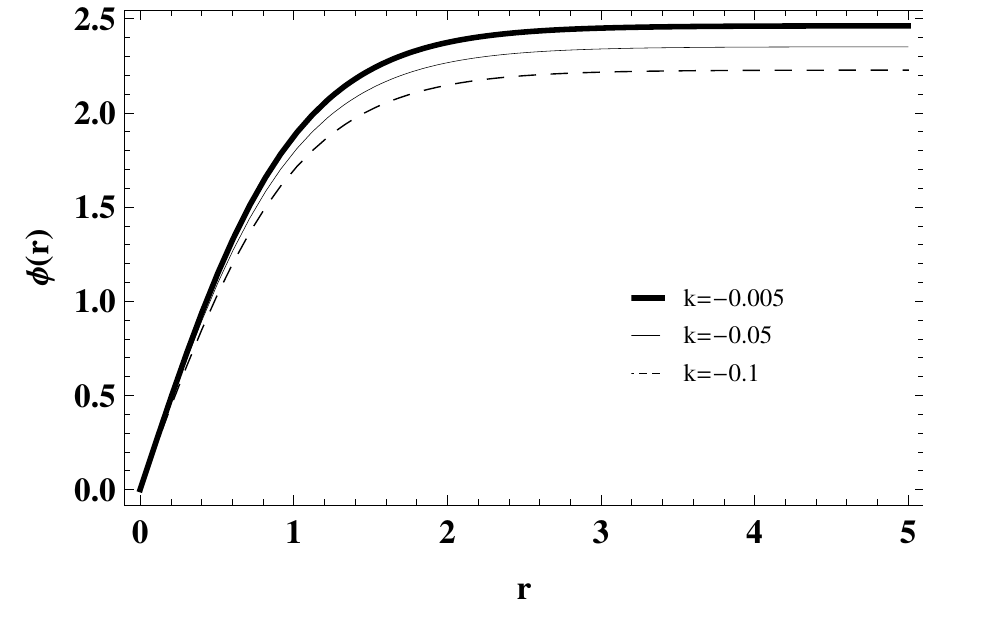} 
\includegraphics[height=5cm]{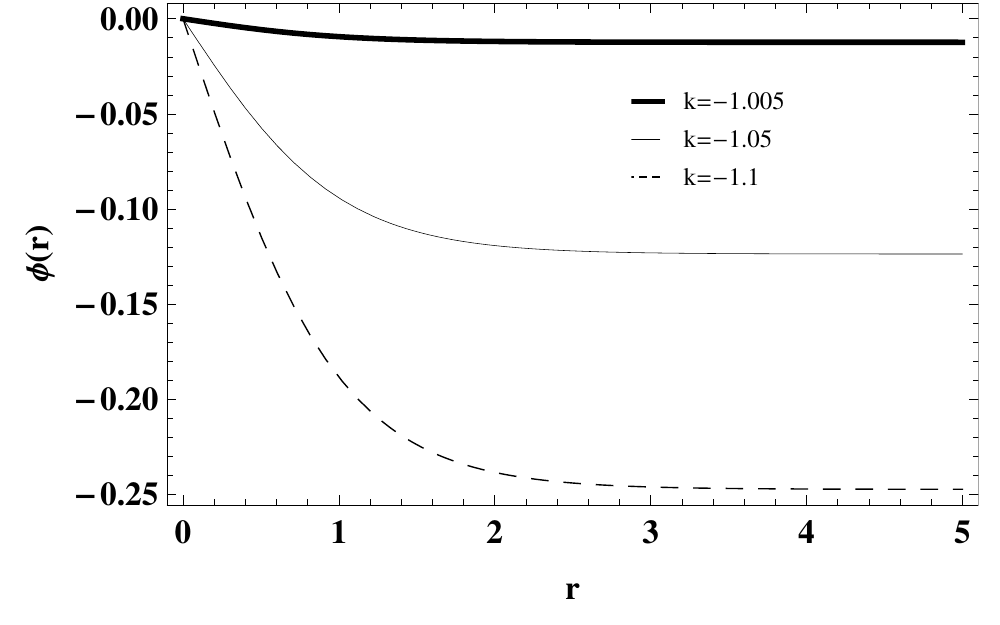}\\
(a) \hspace{8 cm}(b)
\end{tabular}
\end{center}
\caption{ Scalar field $\phi(r)$ for $n=1$, $p=\rho=\lambda=1$. (a) $-1<k<0$ . 
(b) $k<-1$.
\label{figki1}}
\end{figure}

For $n=1$ the solution of Eq.(\ref{q.5}) is given by
\begin{eqnarray}\label{q.7}
\phi(r)=\frac{7}{2\sqrt{2}}(1+k)\tanh(r),
\end{eqnarray}
whose behaviour is depicted in Fig. (\ref{figki1}). Note that for $k>-1$ the solutions resemble kinks whereas for $k<-1$ they behave as anti-kinks.

\begin{figure}
\begin{center}
\begin{tabular}{ccccccccc}
\includegraphics[height=5cm]{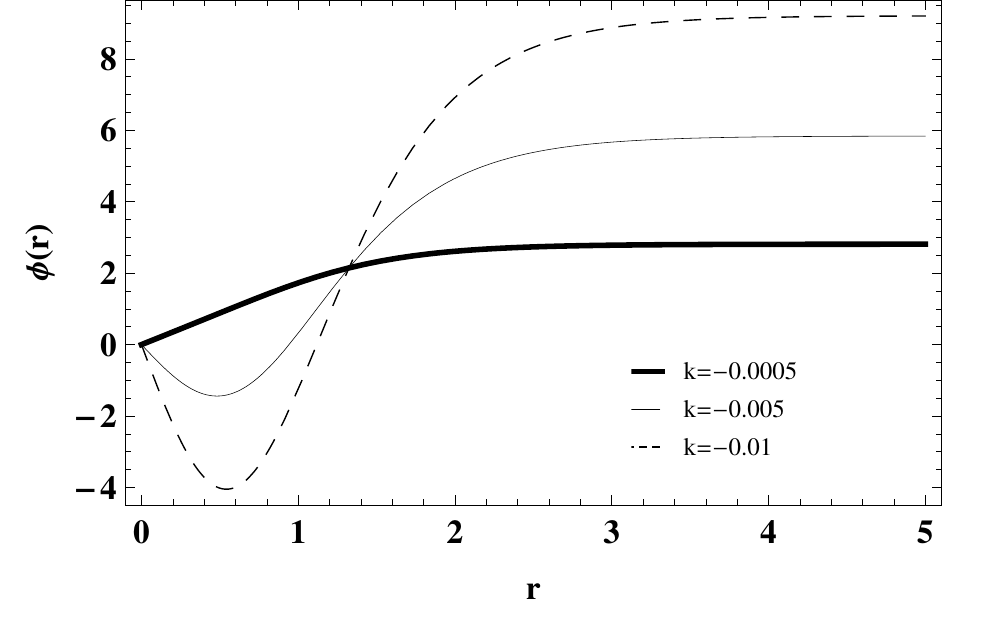} 
\includegraphics[height=5cm]{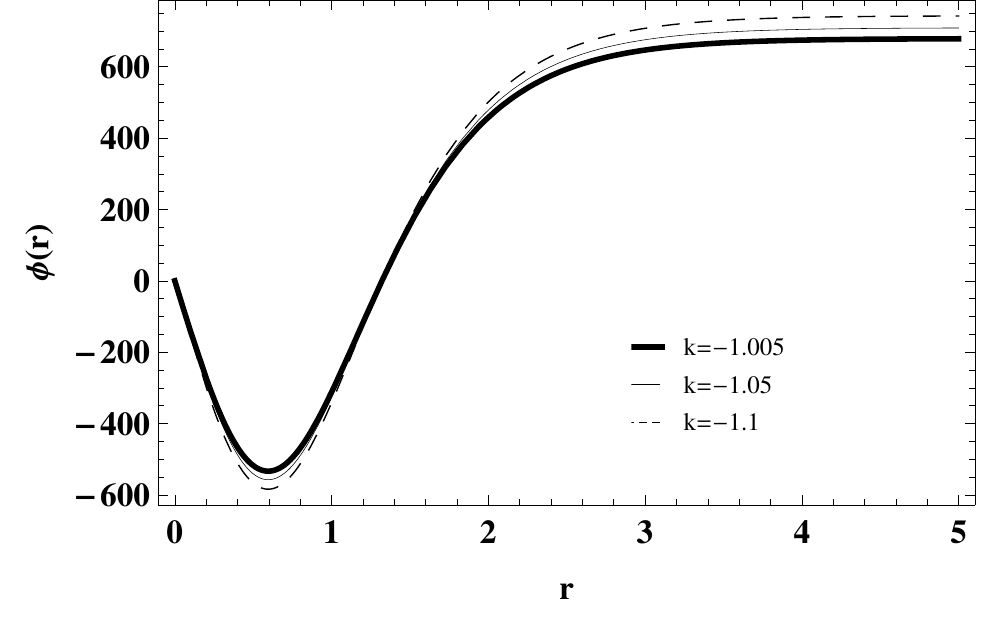}\\
(a) \hspace{8 cm}(b)
\end{tabular}
\end{center}
\caption{Scalar field $\phi(r)$ for $n=3$, $p=\rho=\lambda=1$. (a) $-1<k<0$ . 
(b) $k<-1$.
\label{figki3}}
\end{figure}

For $n=3$,  Eq.(\ref{q.5}) provides the solution
\begin{eqnarray}
\phi(r)=\frac{7}{2\sqrt{2}} \tanh(r)\left[1-272k+1184k\ \mathrm{sech}^2(r)-336k\ \mathrm{sech}^4(r)\right],
\end{eqnarray}
which is sketched in Fig. (\ref{figki3}). Even though the vortex attains the vacuum asymptotically, it exhibits a non monotonic behaviour. This feature suggests that the torsion induces an internal structure for the vortex, likewise the deformation provided by self-interaction potentials or non canonical kinetic terms \cite{ringlike}. Notably, such internal structure matches with the non-monotonic behaviour of the torsion invariant.


\subsubsection{Source properties}

In order to probe further the string-like brane core, we study the distribution of the stress energy tensor components. From 
the angular modified Einstein Eq.(\ref{e.2}) we find the angular pressure in the form 
\begin{eqnarray}
t_\theta(r)&=&-\Lambda + \frac{1}{\kappa_g}\Big\{ g^2 - 4^{n-1}k\left(ug\right)^n -2gy
-2\left[1+4^{n-1}nk\left(ug\right)^{n-1}\right] \left[g\left(1-y\right)-w \right]\Big\}\nonumber\\ 
&-&\frac{4^n k n(n-1)\rho\  \coth(\lambda r)\left(ug\right)^n}{ \lambda\ u^2p \kappa_g} \Bigg\{g\rho\left[\mathrm{csch}^2(\rho r)+ \mathrm{sech}^2(\rho r)+5w\right]\nonumber\\
&-&w\ \mathrm{csch}(\rho r)\ \mathrm{sech}(\rho r)\Bigg\},
\end{eqnarray}
where we defined the functions $g=p \lambda\ \tanh(\lambda r)$, $w= p \lambda^2\ \mathrm{sech}^2( \lambda r)$, $u=4 \rho\ \mathrm{csch}(2\rho r) -5g$ and $y= 2\rho\ \mathrm{csch}(2\rho r) - g$. The angular pressure vanishes asymptotically provided that
\begin{eqnarray}\label{e.005}
\Lambda=-\frac{1}{\kappa_g}\left\{5p^2\lambda^2+(-4)^{n-1}k(2n-1)\left(5p^2\lambda^2\right)^{n}\right\}.
\end{eqnarray}
The fine-tunning Eq. (\ref{e.005}) reveals the relation among the brane width $\epsilon\approx 1/\lambda$, the torsion parameters and the bulk cosmological constant. For $p=1$ and $n=1$, the Eq. (\ref{e.005}) yields to $\lambda=\sqrt{\kappa_g \frac{-\Lambda}{5(k+1)}}=c$. Similar results can be found for $n=2$ and $n=3$. Hence, likewise the 5D $f(T)$ models, the brane width is fine tuned by the bulk cosmological constant \cite{Yang2012}. Note that the higher the $k$ the thicker the brane. Furthermore, for $n=1$ as $k\rightarrow -1$ the brane width tends to $\epsilon\rightarrow 0$. For $\Lambda=0$, the Eq. (\ref{cosmologicalconstant1}) shifts the discontinuity into $k=0$. Therefore, along the $AdS-dS$ transition driven by $k$, the string-like brane undergoes a discontinuous transition between a thick and a thin brane.

Similarly, the energy density and the radial pressure have the form

\begin{eqnarray}
t_0(r)&=&-\Lambda + \frac{1}{\kappa_g}\Big[3 g^2 - 4^{n-1}k\left(ug\right)^n -2g y\Big] +\Bigg[\frac{4^{n-1} k n(n-1)\rho\  \coth^2(\lambda r)\left(ug\right)^n}{\lambda^2\ u^2 p^2\kappa_g} 2y\Bigg]\nonumber\\
&\times& \Big\{g\rho\left[\mathrm{csch}^2(\rho r)+ \mathrm{sech}^2(\rho r)+5w\right]-w\ \mathrm{csch}(\rho r)\ \mathrm{sech}(\rho r)\Big\}\nonumber\\
&+&\frac{1}{2\kappa_g}\left[1+4^{n-1}nk\left(ug\right)^{n-1}\right]y\Big\{2g-4y+4w+\rho^2\left[\mathrm{csch}^2(\rho r)+\mathrm{sech}^2(\rho r)\right]\Big\},
\end{eqnarray}

and 
\begin{eqnarray}
t_r(r)=-\Lambda  -\frac{1}{\kappa_g}\Big[ 3 g^2 - 4^{n-1}(2n-1)k\left(ug\right)^n +2g y\Big].
\end{eqnarray}

\begin{figure}
\begin{center}
\begin{tabular}{ccc}
\includegraphics[height=5cm]{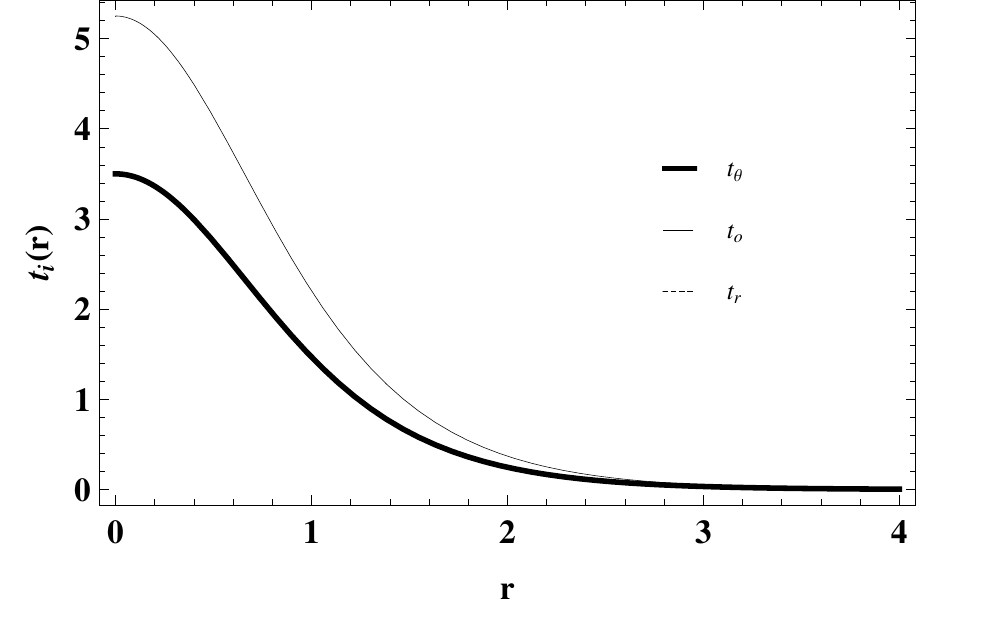}\\ 
(a)\\
\includegraphics[height=5cm]{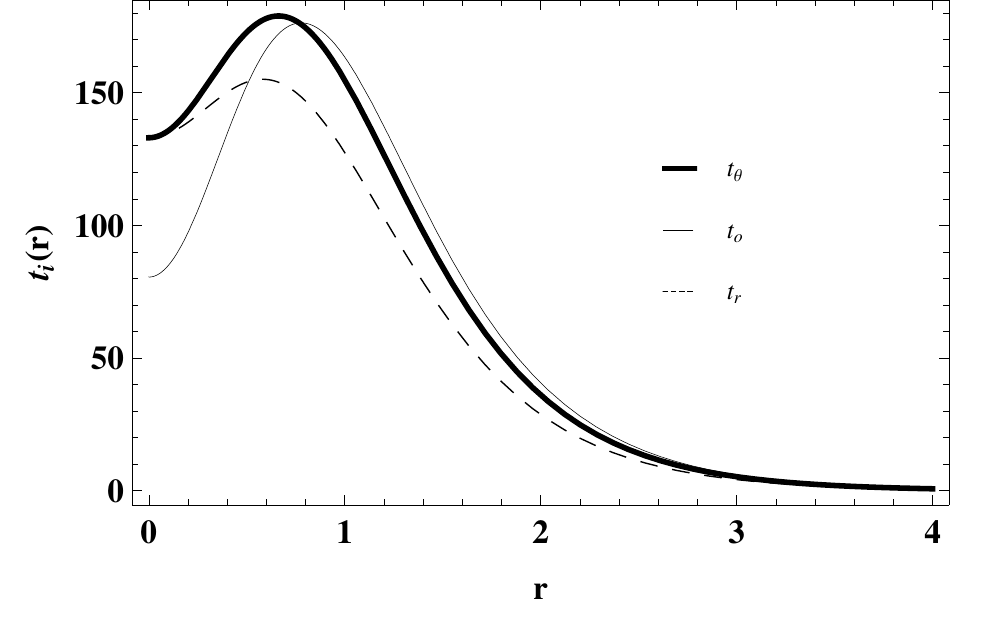} 
\includegraphics[height=5cm]{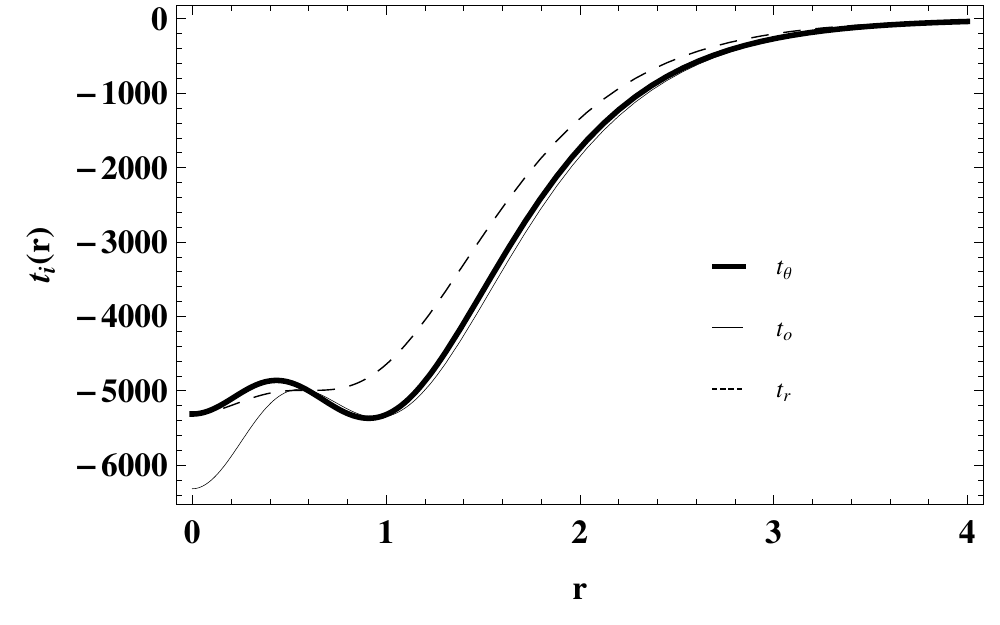}\\
(b) \hspace{8 cm}(c)
\end{tabular}
\end{center}
\caption{ Stress-energy components for  $p=1$, $\rho=1$, $\lambda=1$ and $k=-0.5$. (a) $n=1$. (b) $n=2$. (c) $n=3$.
\label{figene1}}
\end{figure}

In Fig. (\ref{figene1}), we plotted the stress energy components for the same value of $k=-0.5$ and varying the parameter $n$. For $n=1$ (figure $a$), the source exhibits a localized profile satisfying the dominant and strong energy conditions. The $n=2$ configuration (figure $b$) includes a new peak displayed from the origin. That feature reflects the brane internal structure, which tends to split the brane. A similar result was obtained in a codimension one model for a real scalar field \cite{Yang2012}. 
A 6D string-like exhibiting a displayed maximum of the energy density was found in a local vortex with a complex scalar and a gauge field \cite{Giovannini:2001hh}. In a $(2+1)$ flat spacetime, a local ringlike vortex was obtained exhibiting a similar energy density \cite{ringlike}.

A noteworthy feature is the violation of the dominant energy condition for $n=2$. For $n=3$ (figure $c$), the weak energy condition is violated as well. For $n=1$ and high values of $k<0$ or $n=2$ and $0<k<1$, the source presents a negative energy density phase. Therefore, the torsion produces modifications in the source equation of state that might lead to the brane splitting.

\begin{figure}
\begin{center}
\begin{tabular}{ccc}
\includegraphics[height=5cm]{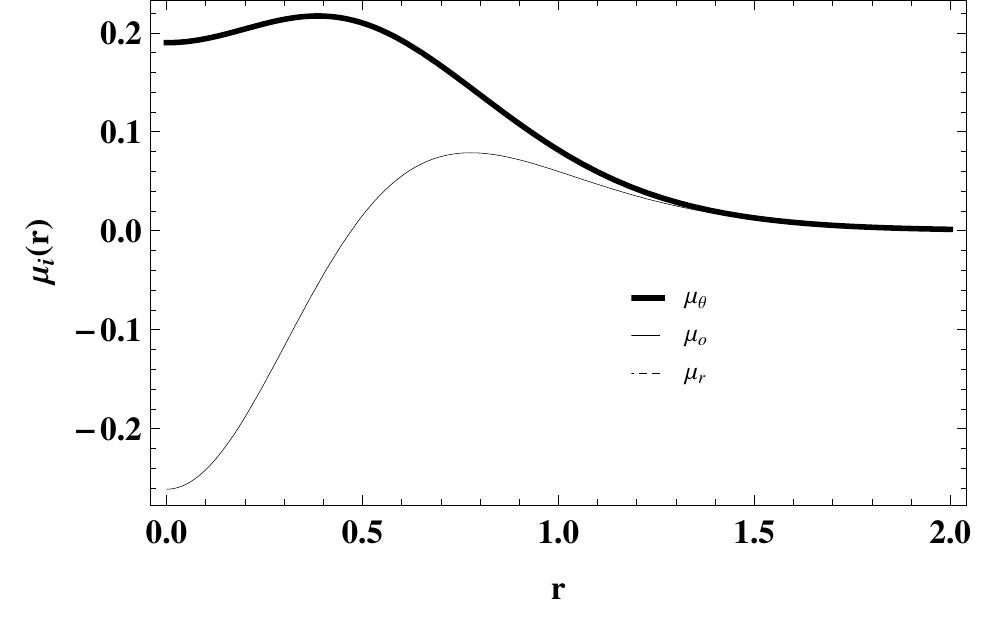} 
\includegraphics[height=5cm]{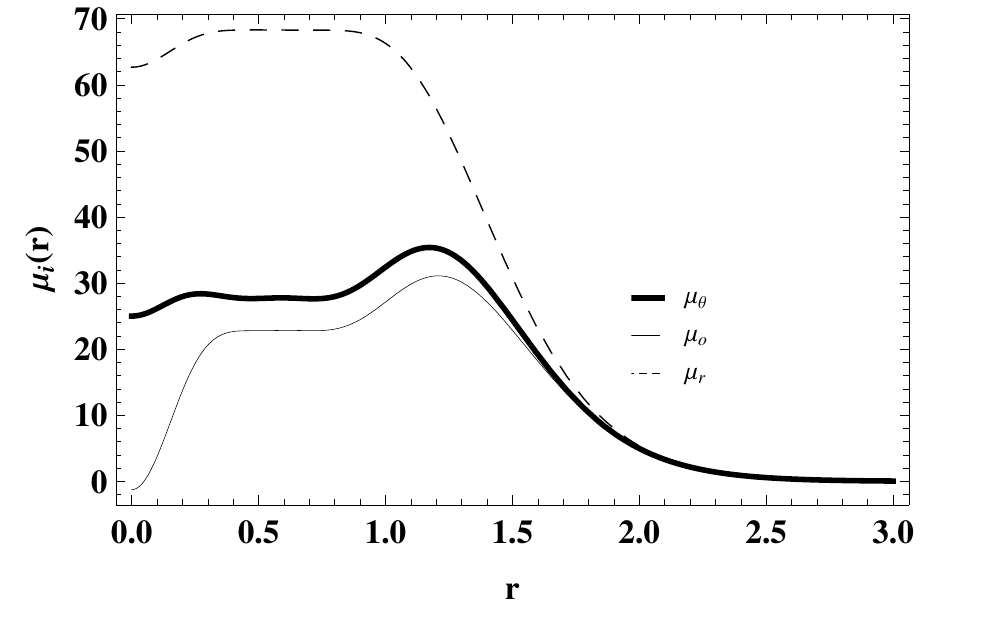}\\
(a) \hspace{8 cm}(b)
\end{tabular}
\end{center}
\caption{ Brane tensions for  $\rho=1$, $\lambda=1$, $p=1$ and $k=-0.05$. (a) $n=1$. (b) $n=3$.} 
\label{branetensions}
\end{figure}

Another important property of the source is the brane tension. The $f(T)$ dynamics modifies the string-like brane tension defined in Ref. \cite{Gherghetta} by
\begin{equation}
\mu_{i}=\int_{0}^{\epsilon}{f_T e^{2A+B}t_{i}(r)dr},
\end{equation}
where $\epsilon$ is the brane width. The Fig. (\ref{branetensions}) shows the variation of the brane tensions inside the brane core. For $n=1$ the angular and radial tensions coincide whereas the timelike tension is negative. The $n=3$ configuration reveals positive brane tensions with a plateau inside the core. Unlike the GS model, the $\mu_0$ tension is lesser than $\mu_r$ and $\mu_\theta$. 

\section{Gravitational Perturbations}
\label{sec4}

In this section we investigate the effects of torsion on the propagation of the gravitational perturbations.  
We follow closely the analysis performed in Ref.\cite{tensorperturbations} and extend it to a six dimensional bulk.

Consider the \textit{seschsbein} perturbation
\begin{displaymath}
h^a\ _\mu=\left(\begin{array}{cccccc}
e^{A(r)}\left(\delta^a_\mu+w^a\ _\mu\right)&0&0\\
0&1&0\\
0&0&R_0e^{B(r)}\\
\end{array}\right),
\end{displaymath}
where $w^a\ _\mu=w^a\ _\mu(x^\mu,r,\theta)$. The resulting metric perturbation takes the form $ds^2=e^{A(r)}\left(\eta_{\mu\nu}+\gamma_{\mu\nu}\right)dx^\mu dx^\nu+dr^2+R^2_0e^{2B(r)}d\theta^2$,
where the metric and the \textit{sechsbein} perturbation are related by
\begin{eqnarray}
 \gamma_{\mu\nu}&=&(\delta^a_\mu w^b\ _\nu+\delta^b_\nu w^a\  _\mu)\eta_{ab},\nonumber\\
 \gamma^{\mu\nu}&=&(\delta_a^\mu w_b\ ^\nu+\delta_b^\nu w_a\  ^\mu)\eta^{ab}.
\end{eqnarray}
Assuming the transverse traceless metric gauge $\partial_\mu \gamma^{\mu\nu}=0=\eta^{\mu\nu}\gamma_{\mu\nu}$ 
leads to the \textit{sechsbein} gauge 
\begin{equation}
\delta_a^\mu w^a\  ^\mu=0,
\end{equation}
The non-vanishing components of the torsion tensor are
\begin{eqnarray}
\label{23.l}
 T^\rho\ _{\mu r}&=&-A'\delta^\rho_\mu-(\delta^\rho_a w^a\ _\mu-\delta^a_\mu w_a\ ^\rho)A'-\delta^\rho_a w^a\ _\mu, \nonumber \\
 T^\rho\ _{\mu \nu}&=&\delta^\rho_a(\partial_\mu w^a\ _\nu-\partial_\nu w^a\ _\mu), \nonumber \\
 T^\rho\ _{\mu \theta}&=&-\delta^\rho_a\partial_\theta w^a\ _\mu, \nonumber \\
 T^\theta\ _{\theta r}&=& -B',
\end{eqnarray}
whereas the non-vanishing contorsion components are

\begin{eqnarray}\label{24.l}
 K^\rho\ _{\mu r}&=&A'(\delta^\rho_a w^a\ _\mu-\delta^a_\mu w_a\ ^\rho)+\frac{1}{2}\left(\delta^\rho_a w'^a\ _\mu-\delta^a_\mu w'_a\ ^\rho\right),\nonumber\\
 K^\rho\ _{r \nu}&=&-A'\delta^\rho_\nu-\frac{1}{2}\left(\delta^a_\nu w'_a\ ^\rho-\delta_a^\rho w'^a\ _\nu\right),\nonumber\\
 K^r\ _{\mu \nu}&=&e^{2A}(A'\eta_{\mu\nu}+A'\gamma_{\mu\nu}+\frac{1}{2}\gamma'_{\mu\nu}),\nonumber\\
 K^\rho\ _{\mu \nu}&=& \frac{1}{2}\left[\delta^a_\mu(\partial^\rho w_{a\nu}-\partial_\nu w_a\ ^\rho)+\delta^a_\nu(\partial^\rho w_{a\mu}-\partial_\mu w_a\ ^\rho)-\delta_a^\rho(\partial_\mu w^a\ _{\nu}-\partial_\nu w^a\ _\mu)\right],\nonumber\\
 K^\rho\ _{\mu \theta}&=&\frac{1}{2}\left(\delta_a^\rho\partial_\theta w^a\ _{\mu}-\delta^a_\mu\partial_\theta w_a\ ^\rho\right),\nonumber\\
 K^\rho\ _{ \theta\nu}&=&-\frac{1}{2}\left(\delta_a^\rho\partial_\theta w^a\ _{\nu}+\delta^a_\nu\partial_\theta w^a\ _\mu\right),\nonumber\\
 K^\theta\ _{ \mu\nu}&=&\frac{1}{2}\left(\delta_{a\mu}\partial_\theta w^a\ _{\nu}+\delta^{a\nu}\partial_\theta w_a\ ^\rho\right)e^{2A}g^{\theta\theta},\nonumber\\
 K^\theta\ _{r \theta}&=& -B' ,\nonumber\\
 K^r\ _{\theta \theta}&=& B'g_{\theta\theta}.
\end{eqnarray}
Accordingly, the non-vanishing components of the dual torsion tensor are
\begin{eqnarray}\label{25.l}
 S_\rho\ ^{\mu r}&=&\frac{1}{2}\left[(3A'+B')\delta^\mu_\rho-\frac{1}{2}(\delta_\rho^a w'_a\ ^\mu+\delta_a^\mu w'^a\ _\rho)\right],\nonumber\\
 S_r\ ^{\mu \nu}&=&\frac{1}{2}\left[A'(\delta^\mu_a w^{a\nu}-\delta_a^\nu w^{a\mu})+\frac{1}{2}(\delta^\mu_a w'^{a\nu}-\delta_a^\nu w'^{a\mu})\right]e^{-2A},\nonumber\\
 S_\rho\ ^{\mu \nu}&=&\frac{1}{4}\left[\delta^\nu_a(\partial^\mu w^{a}\ _\rho-\partial_\rho w^{a\mu})-\delta^\mu_a(\partial^\nu w^{a}\ _\rho-\partial_\rho w^{a\nu})\right]e^{-2A}+\frac{1}{4}\delta_\mu^a(\partial^\mu w_{a}\ ^\nu-\partial^\nu w_a\ ^{\mu})\nonumber\\ &+&
\frac{1}{2}\left[\delta^\nu_\rho\delta^\lambda_a\partial_\lambda w^{a\mu}-\delta^\mu_\rho\delta^\lambda_a\partial_\lambda w^{a\nu}\right]e^{-2A},\nonumber\\
 S_r\ ^{\mu r}&=&\frac{1}{2}(\delta^\rho_a\partial_\rho w^{a\mu})e^{-2A},\nonumber\\
 S_\rho\ ^{\mu \theta}&=&-\frac{1}{4}(\delta_\rho^a\partial_\theta w_{a}\ ^\mu+\delta^\mu_a\partial_\theta w^a\ _\rho)g^{\theta\theta},\nonumber\\
 S_\theta\ ^{\mu \nu}&=&\frac{1}{4}(\delta^\mu_a\partial_\theta w^{a\nu}+\delta^{a\nu}\partial_\theta w_a\ ^\mu)e^{-2A},\nonumber\\
 S_\theta\ ^{\theta r}&=&-2A'.
\end{eqnarray}
The linearlized modified Einstein equation Eq.(\ref{3.36}) has the form
\begin{eqnarray}\label{27.l}
\frac{1}{h}f_T\Bigg[\delta g_{NP}\partial_Q\left(h S_M\ ^{PQ}\right)+ g_{NP}\partial_Q\left(h \delta S_M\ ^{PQ}\right)&\nonumber\\
-h\left(\delta\widetilde{\Gamma}^Q\ _{PM}S_{QN}\ ^{P}+h\widetilde{\Gamma}^Q\ _{PM}\delta S_{QN}\ ^{P}\right)\Bigg] +f_{TT}\delta S_{MN}\ ^{Q}\partial_Q T+\frac{1}{4}\delta g_{MN}f=&\delta\mathcal{T}_{MN},
\end{eqnarray}
which for $\delta h=0$ and $\delta T=0$ yields to
\begin{eqnarray}\label{28.l}
-\frac{1}{4}\left[e^{-2A}\Box \gamma_{\mu\nu}+(4A'+B')\gamma'_{\mu\nu}+\gamma''_{\mu\nu}+R_0^{-2}e^{-2B}\partial_\theta^2\gamma_{\mu\nu}\right]e^{2A}f_T &\nonumber\\
+\frac{1}{2}\left[(4A'+B')(3A'+B')+3A''+B''\right]\gamma_{\mu\nu}e^{2A}f_T+\frac{1}{4}\gamma_{\mu\nu}e^{2A}f &\nonumber\\
 -f_{TT} \left[2(3A'+B')\gamma_{\mu\nu}-\gamma'_{\mu\nu}\right]\left[A''(3A'+2B')+A'(3A''+2B'')\right]e^{2A}=&\delta\mathcal{T}_{\mu\nu},
\end{eqnarray}
where $\Box=\eta^{\mu\nu}\partial_\mu \partial_\nu$ and the radial and angular perturbations vanish. The linearlized stress energy tensor writes 
\begin{eqnarray}\label{29.l}
\delta\mathcal{T}_{\mu\nu}=\delta(\mathcal{T}_{\mu}\ ^\mu g_{\mu\nu})=\delta(\mathcal{T}_{\mu}\ ^\mu)\eta_{\mu\nu}e^{2A}+\mathcal{T}_{\mu}\ ^\mu \gamma_{\mu\nu}e^{2A}.
\end{eqnarray}
The gravitational field equation provides the condition
\begin{eqnarray}\label{30.l}
\frac{1}{2}\left[(4A'+B')(3A'+B')+3A''+B''\right]f_T &\nonumber\\ - 2(3A'+B')\left[A''(3A'+2B')+A'(3A''+2B'')\right]f_{TT} +\frac{1}{4}f=&\mathcal{T}_{\mu}\  ^\mu.
\end{eqnarray}
Employing the condition (\ref{30.l}) and the vanishing trace $\delta(\mathcal{T}_{\mu}\ ^\mu)$ gives the perturbation equation
\begin{eqnarray}\label{32.l}
\left[e^{-2A}\Box \gamma_{\mu\nu}+(4A'+B')\gamma'_{\mu\nu}+\gamma''_{\mu\nu}+R_0^{-2}e^{-2B}\partial_\theta^2\gamma_{\mu\nu}\right]f_T&\nonumber\\ -4 \left[A''(3A'+2B')+A'(3A''+2B'')\right]\gamma'_{\mu\nu}f_{TT}=&0.
\end{eqnarray}
Assuming the Kaluza-Klein decomposition $\gamma_{\mu\nu}(x^\rho,r,\theta)=\epsilon_{\mu\nu}(x^\rho)\sum_{\beta=1}^{\infty} \chi(r) e^{i\beta\theta}$ and a 4D plane-wave satisfying $\left(\Box-m_0^2\right)\epsilon_{\mu\nu}=0$, the perturbed Einstein equation Eq.(\ref{32.l}) yields to
\begin{eqnarray}
\label{KKequation}
\chi''+\left\{4A'+B'-4 \left[A''(3A'+2B')+A'(3A''+2B'')\right]\frac{f_{TT}}{f_{T}}\right\}\chi'&\nonumber\\
+\left(e^{-2A}m_0^2-R_0^{-2}e^{-2B}\beta^2\right)\chi=&0.
\end{eqnarray}
The torsion adds a new term proportional to $f_{TT}/f_{T}$ when compared to the GR based string-like braneworld \cite{conifold,cigar}.

\subsubsection{Kaluza-Klein modes}
\begin{figure}
\begin{center}
\begin{tabular}{ccccccccc}
\includegraphics[height=5cm]{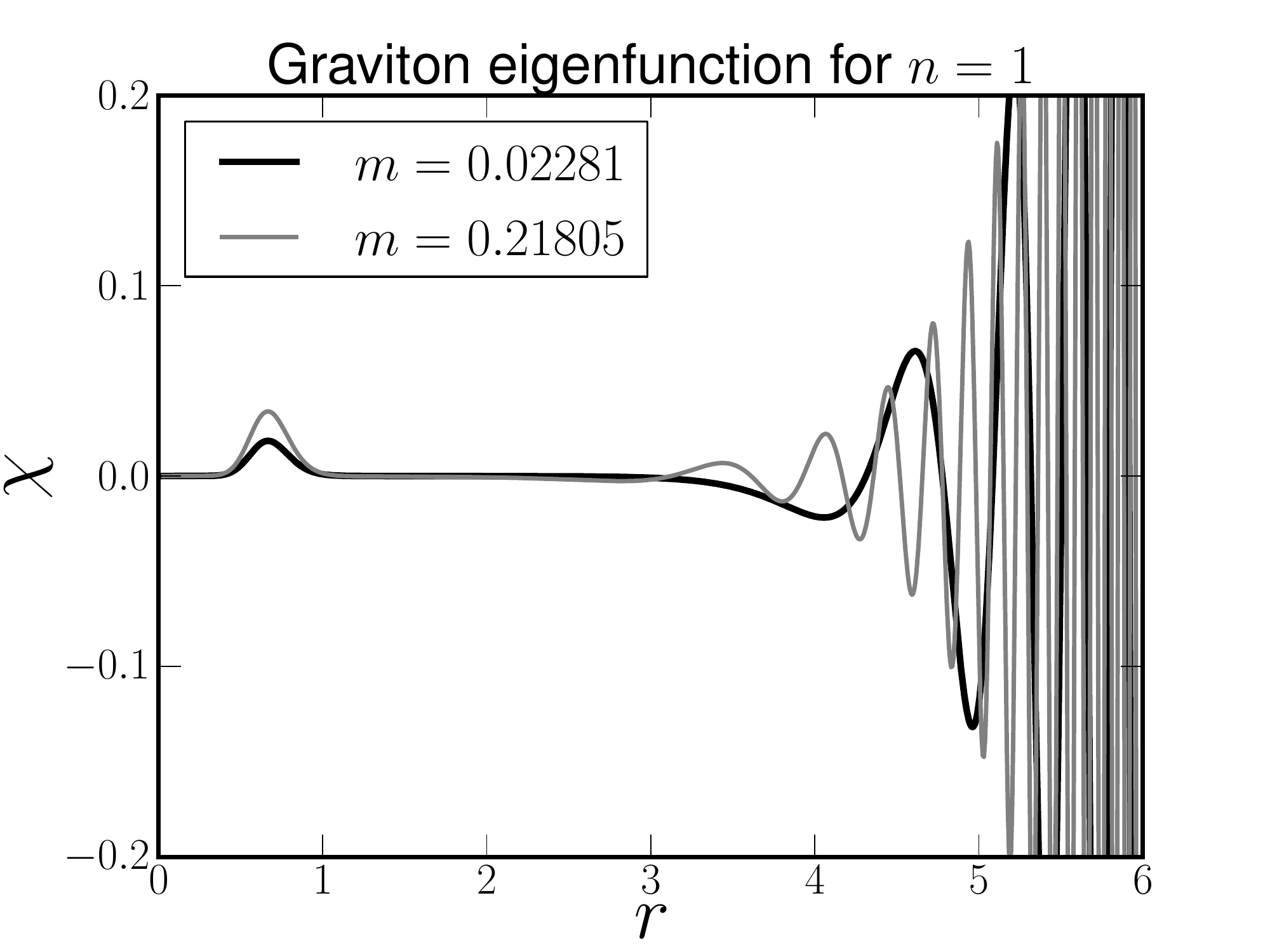} \\
(a)\\
\includegraphics[height=5cm]{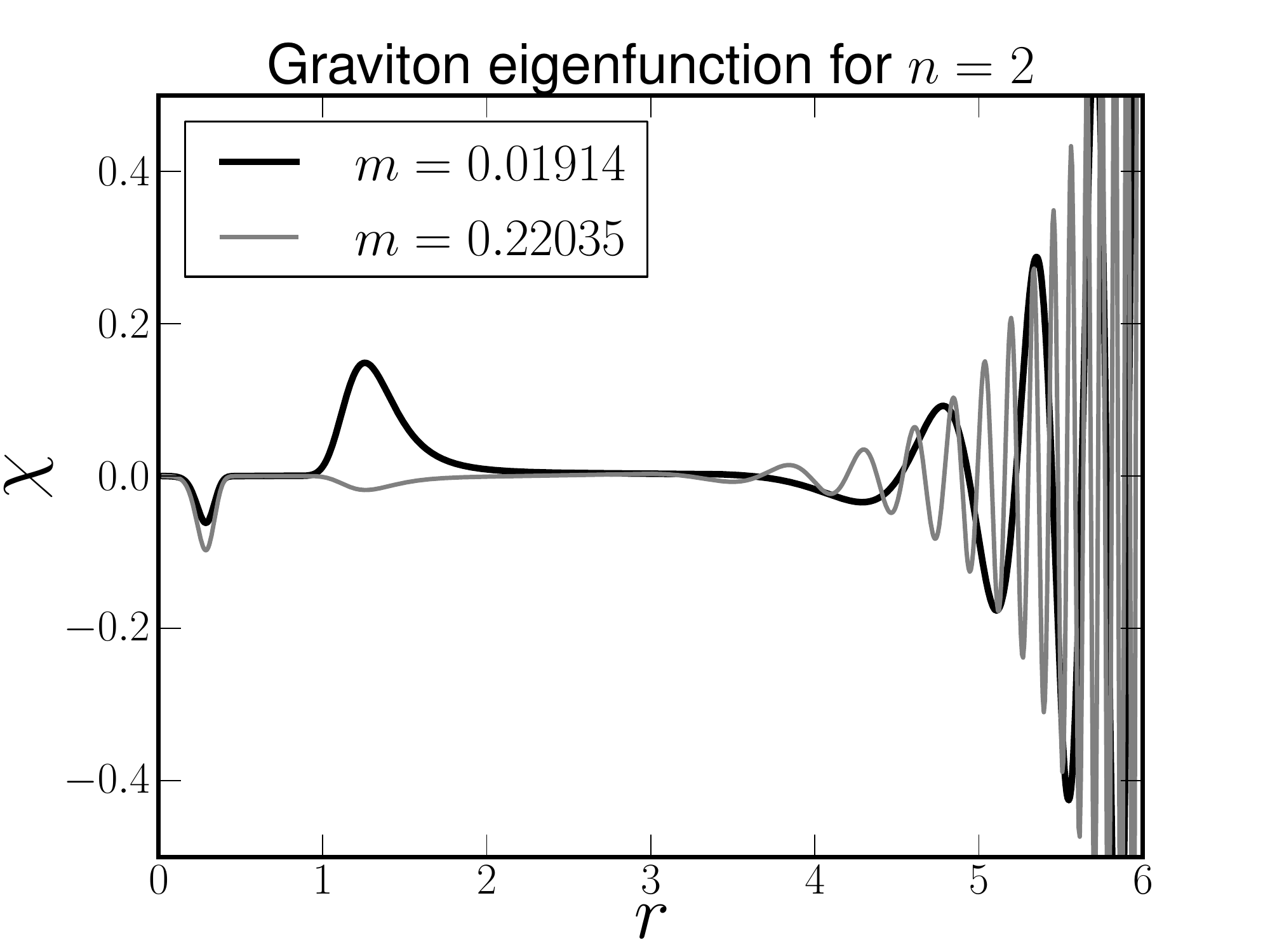} 
\includegraphics[height=5cm]{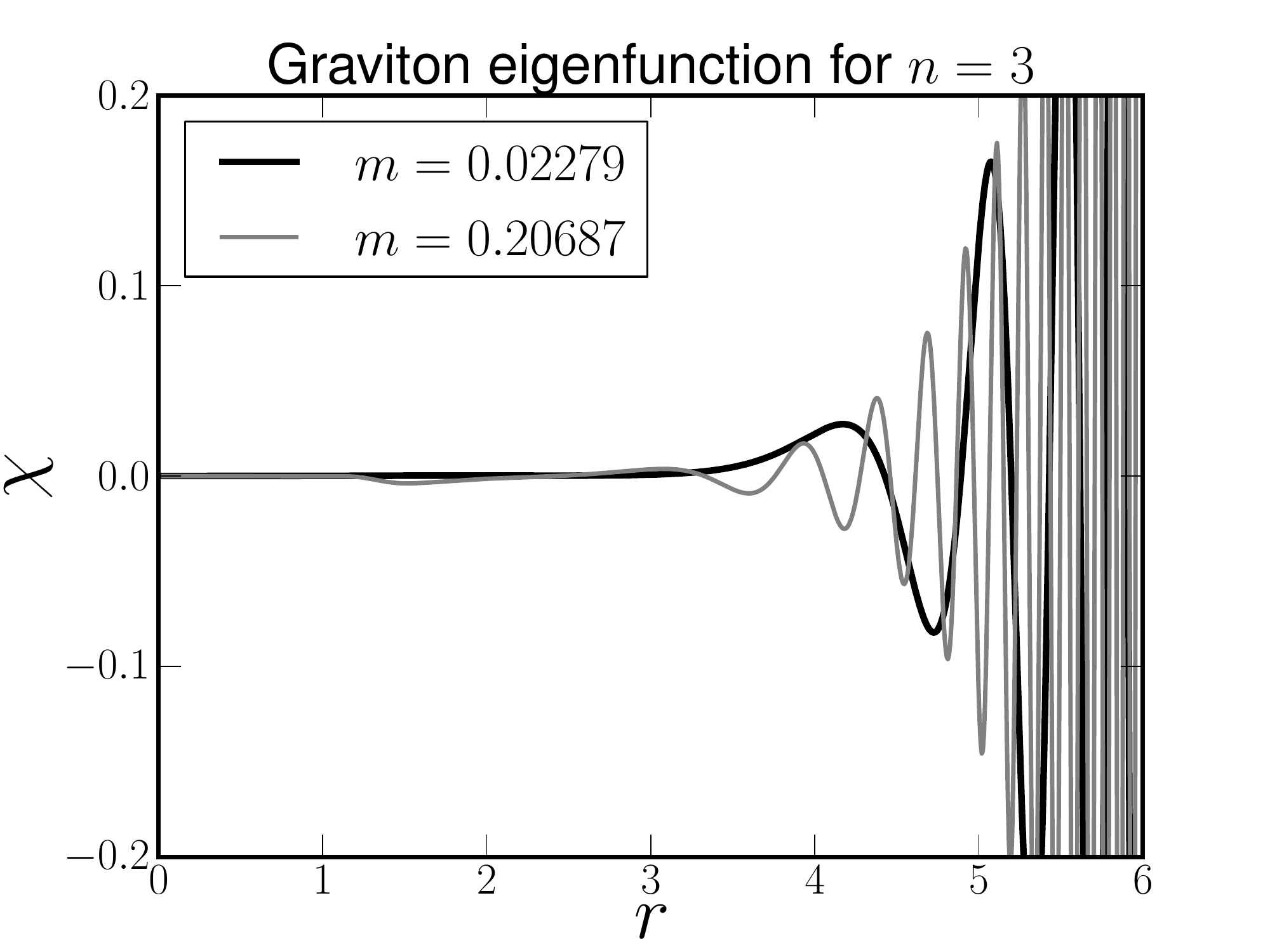}\\
(b) \hspace{8 cm}(c)
\end{tabular}
\end{center}
\caption{Massive modes for $p=\rho=\lambda=1$. (a) $n=1$. (b)  $n=2$. (c) $n=3$.} 
\label{massivemodes}
\end{figure}

Let us firstly consider the effects of torsion in the region exterior to the brane. That limit can also be interpreted as representing a thin string-like brane. In this regime $A'=B'=-c$ and the Eq.(\ref{KKequation}) takes the form
\begin{eqnarray}\label{91.l}
\chi''-5c\chi'+e^{2cr}\left(m_0^2-R_0^{-2}\beta^2\right)\chi=0.
\end{eqnarray}
The Eq.(\ref{91.l}) is the same of the thin string-like brane in GS model \cite{Gherghetta}. As depicted in Fig.(\ref{massivemodes}) the asymptotic divergence of the massless gravitational shows that they form a tower of non-localized states.
Nearby the origin, the torsion modifies the KK modes by
\begin{eqnarray}\label{93.l}
\chi''-\left\{\frac{120r^2+(15kn^28^nr^2+16r^2-24)+8^nkn[2(2n-1)r^2-3]}{3r(8+8^nkn)}\right\}\chi'-m_0^2\chi=0,
\end{eqnarray}
whose finite solutions are given by the Kummer Hypergeometric confluent function
\begin{eqnarray}
\chi(r)=C\ _1F_1 \left(\frac{12m_0^2+3(2)^{3n-1}nkm_0^2}{136-2^{3n+1}nk19(2)^{3n}kn^2},1, \frac{136-2^{3n+1}nk19(2)^{3n}kn^2}{12m_0^2+3(2)^{3n-1}nkm_0^2}r^2\right),
\end{eqnarray}
where $C$ is a constant.

\begin{figure}
\begin{center}
\begin{tabular}{ccccccccc}
\includegraphics[height=5cm]{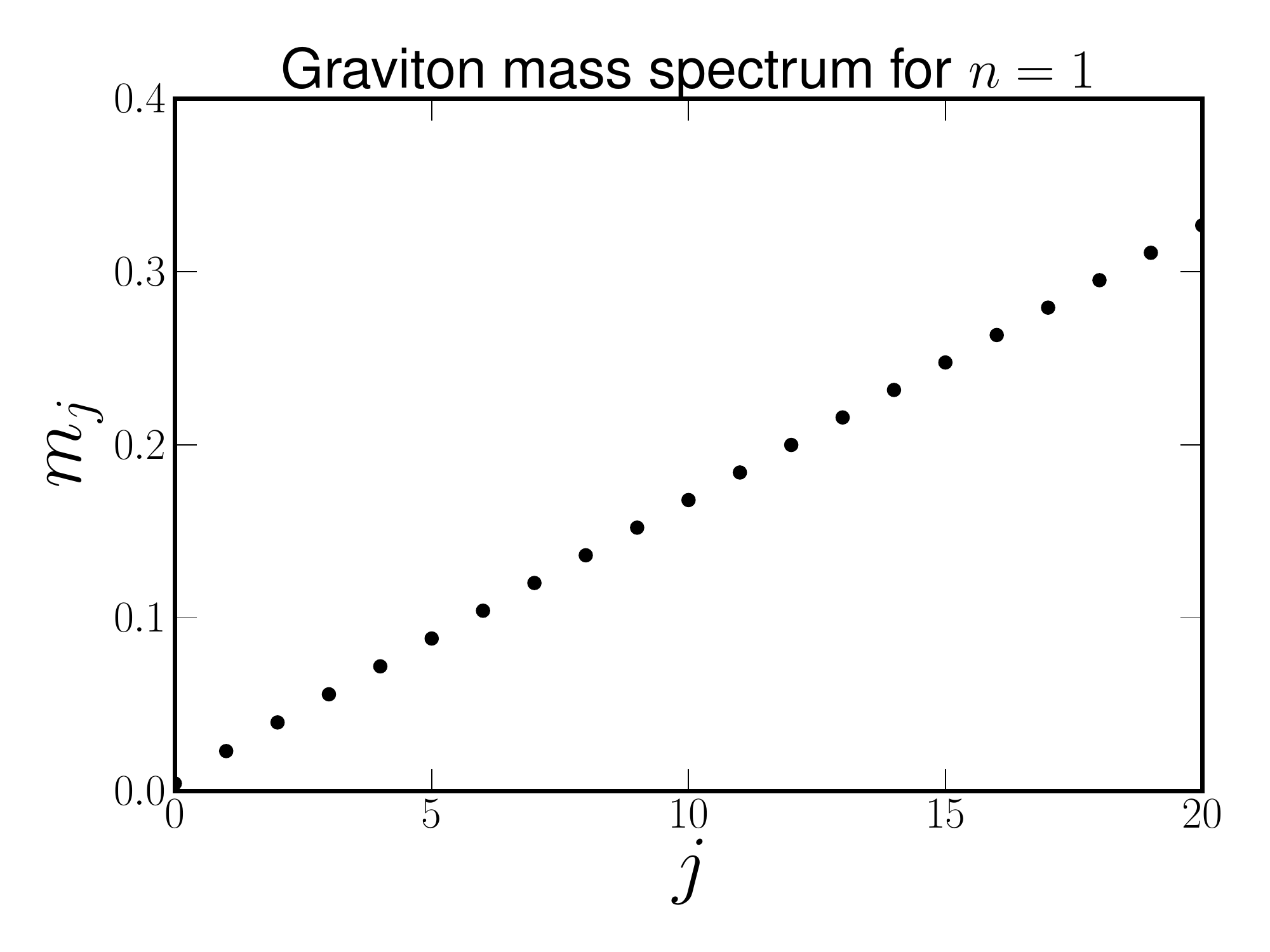} \\
(a)\\
\includegraphics[height=5cm]{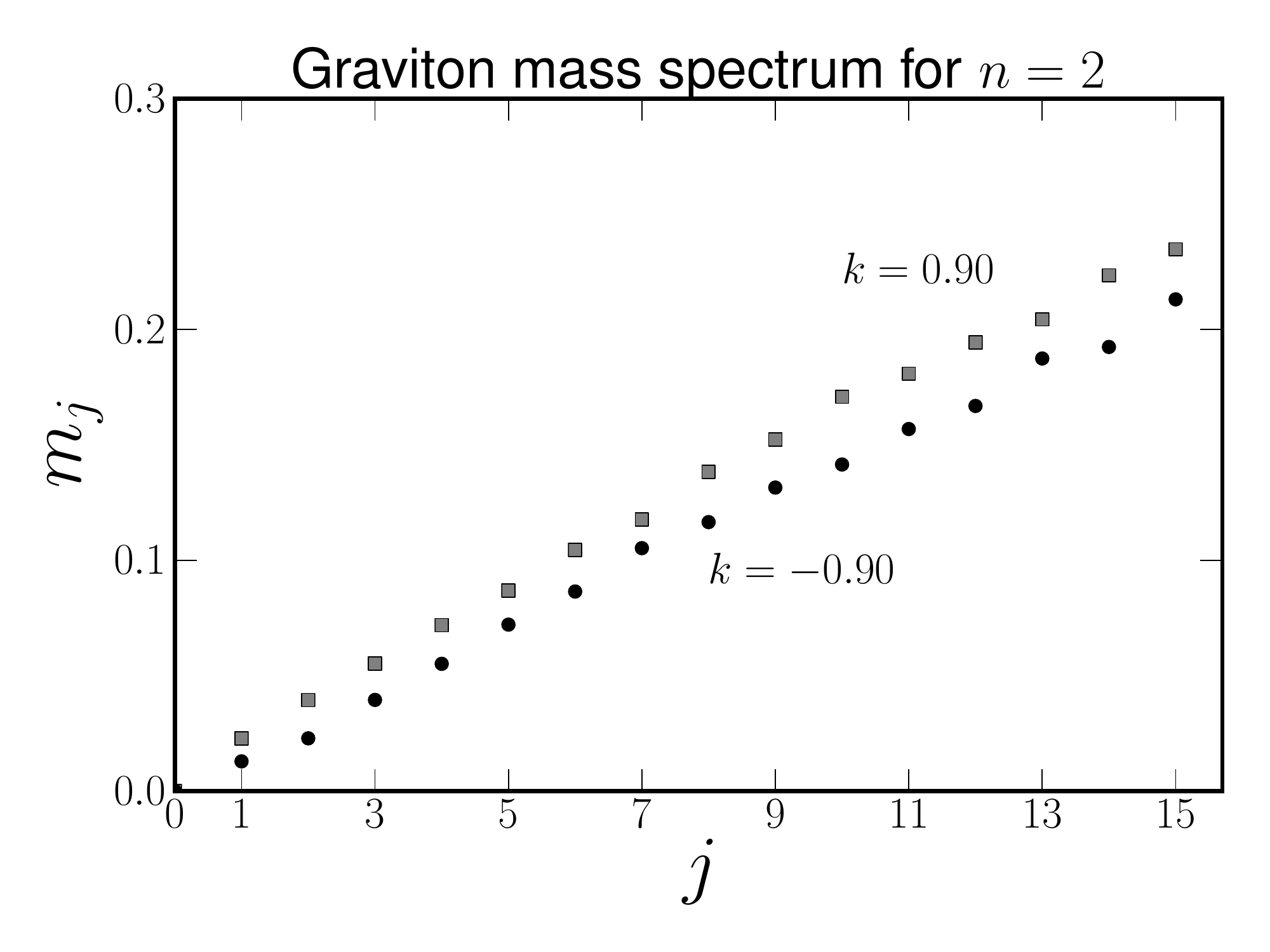} 
\includegraphics[height=5cm]{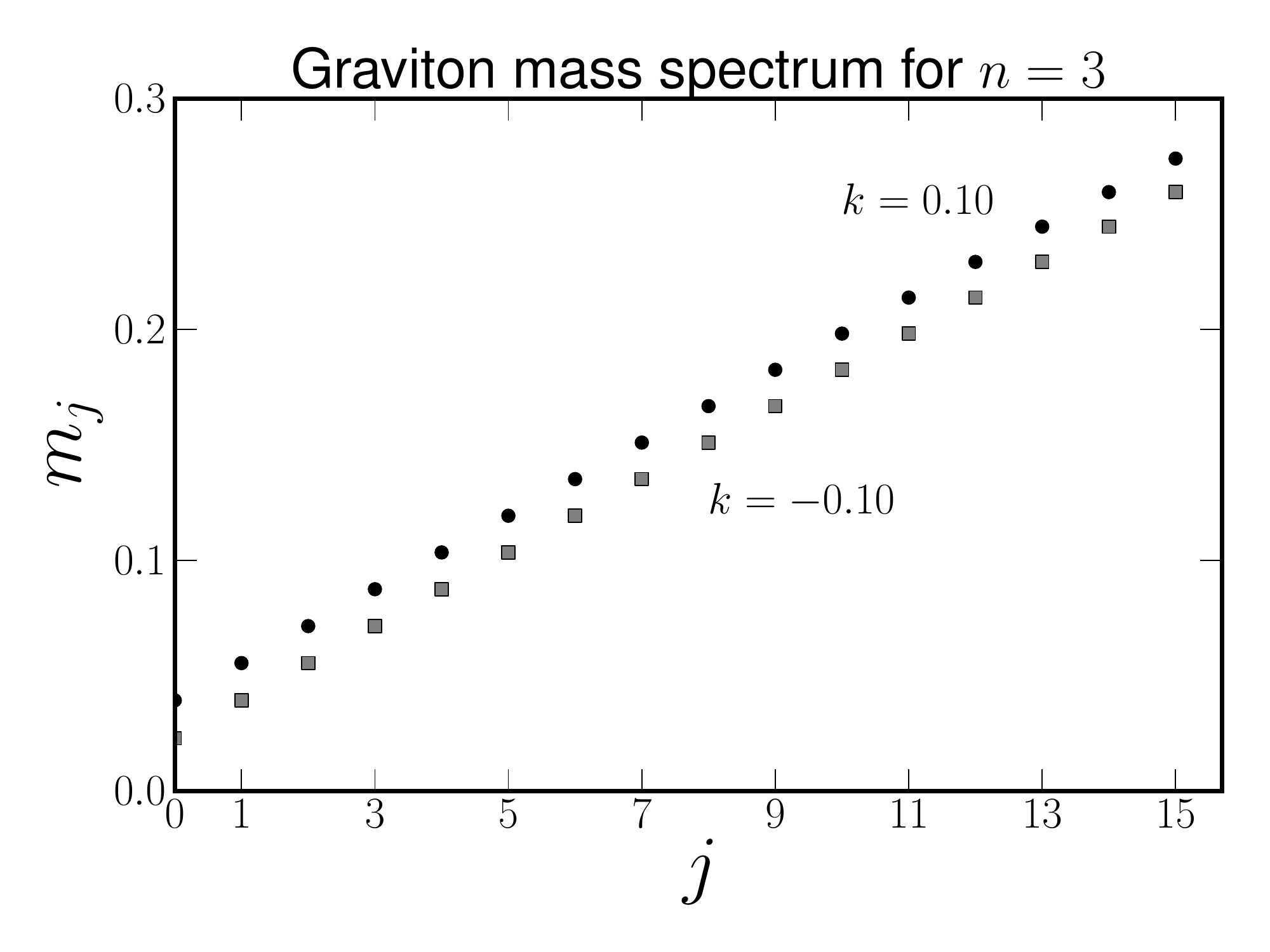}\\
(b) \hspace{8 cm}(c)
\end{tabular}
\end{center}
\caption{Massive spectrum for $p=\rho=\lambda=1$. (a) $n=1$. (b)  $n=2$. (c) $n=3$.} 
\label{spectrum}
\end{figure}

By employing the matrix method, we numerically solved the Eq. (\ref{KKequation}), thereby obtaining the KK spectrum and the respective massive modes. Adopting the usual boundary condition $\chi'(0)=\chi'(\infty)=0$ \cite{Gherghetta,KK1} with a cut off at $r=6$ and $N=1.200$ subdivisions, the first massive modes are shown in Fig. (\ref{massivemodes}), whereas the spectrum is depicted in Fig. (\ref{spectrum}). 

For $n=1$ the massive modes exhibit a single bump inside the thick string-like brane, whereas for $n=2$ two bumps appear. This seems to indicate a splitting process of the massive modes inside the brane as it splits from a thick string-like into ring-like branes. On the other hand, the $n=3$ configuration has no small amplitude inside the brane. We discuss more on this behaviour in the following subsection. 

\begin{figure}
\begin{center}
\begin{tabular}{ccc}
\includegraphics[height=5cm]{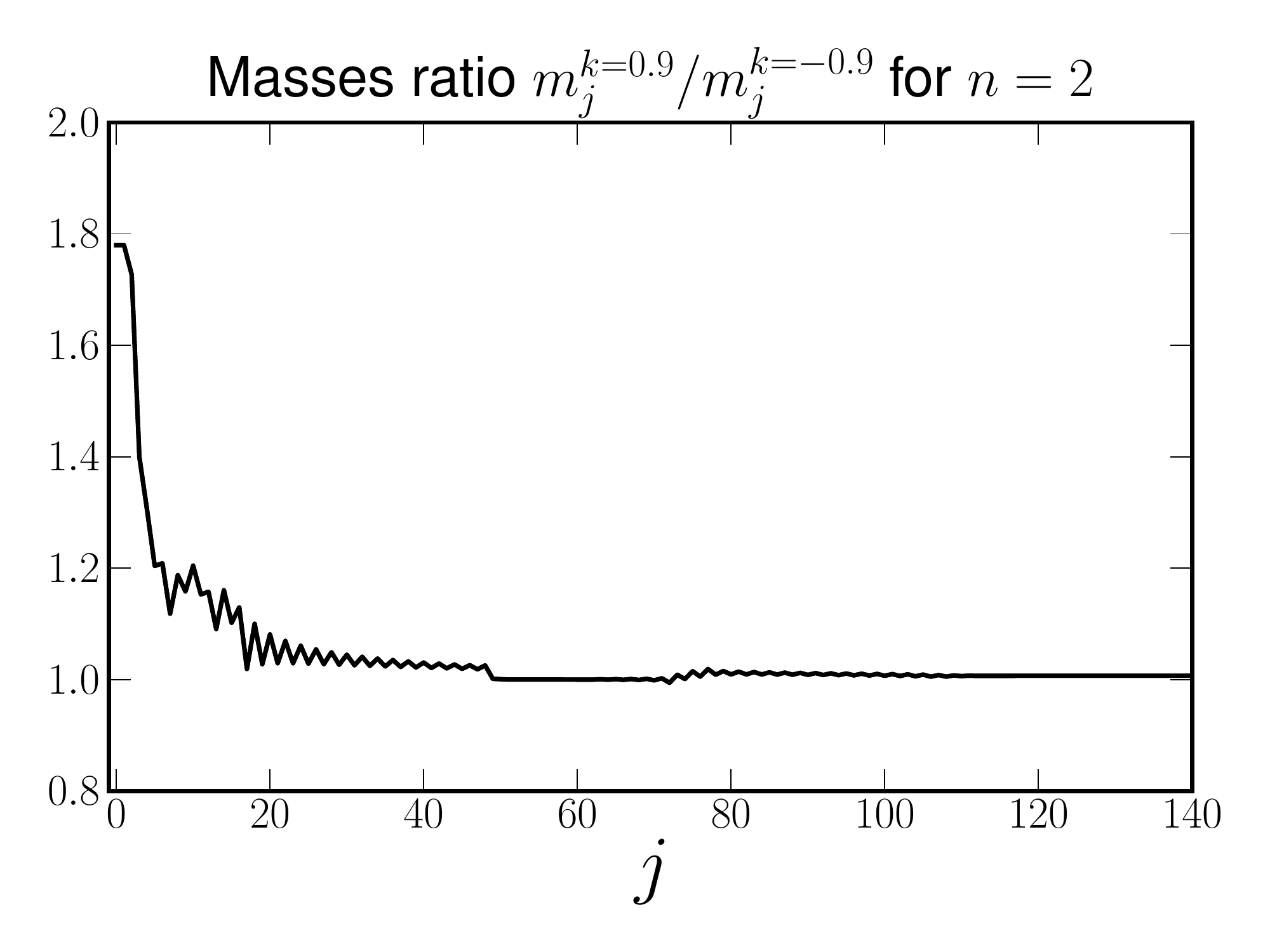} 
\includegraphics[height=5cm]{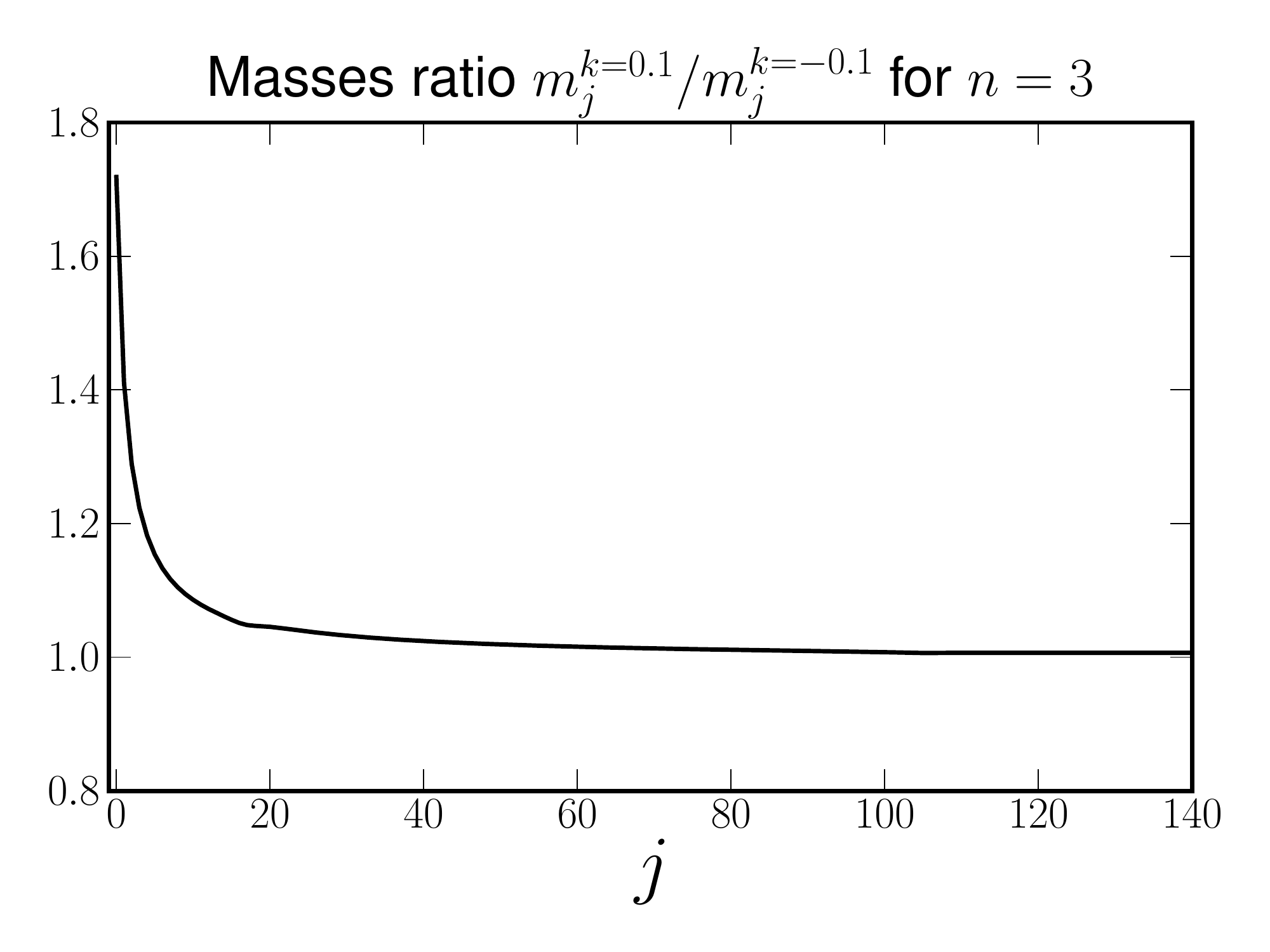}\\
(a) \hspace{8 cm}(b)
\end{tabular}
\end{center}
\caption{Ratio between the KK masses. (a) For $n=2$.
(b) For $n=3$.} 
\label{ratio}
\end{figure}

The spectrum shown in Fig. (\ref{spectrum}) reveals an usual linear behaviour for small masses \cite{Gherghetta,KK1}. For $n=1$ the spectrum is independent of the torsion parameter $k$, as expected from the expression Eq.(\ref{KKequation}). For $n=2$ and $n=3$, by considering opposite values of $k$ the spectrum suffers a shift for the first masses. Nonetheless, as the masses increase, their values tend to be independent of the torsion parameter $k$, as shown in Fig. (\ref{ratio}).

\subsection{Analogue potential}

Employing the change to a conformal coordinate $z=\int{e^{-A}}dr$ and the change on the wave function $\chi(z)=e^{-\frac{1}{2}(3A+B)+\int K(z)dz}\Psi(z)$, the KK Eq. (\ref{KKequation}) can be recast into a Sch\"{o}dinger-like equation 
\begin{eqnarray}\label{36.l}
\left[-\partial_z^2+U(z)\right]\Psi(z)=m^2\Psi(z),
\end{eqnarray}
where the potential is defined by
\begin{eqnarray}\label{potential}
U(z)=\dot{H}+H^2,
\end{eqnarray}
where the prime $(\ \dot{}\ )$  denotes differentiation with respect to $z$,  with
\begin{eqnarray}
\label{kz}
 K(z)=-4e^{-2A}\left[3\left(\dot{A}^3-\dot{A}\ddot{A}\right)+2\dot{A}^2\dot{B}-\dot{B}\ddot{A}-\dot{A}\ddot{B}\right] \frac{f_{TT}}{f_T},
\end{eqnarray}
and $m^2=m_0^2-\beta^2R_0^{-2}e^{2(A-B)}$. The superpotential $H$ is given by 
\begin{eqnarray}\label{34.l}
H=\frac{1}{2}\left(3\dot{A}+\dot{B}\right)+4e^{-2A}\left[3\left(\dot{A}^3-\dot{A}\ddot{A}\right)+2\dot{A}^2\dot{B}-\dot{B}\ddot{A}-\dot{A}\ddot{B}\right] \frac{f_{TT}}{f_T},
\end{eqnarray}
and the quantum mechanic supersymmetric form of the potential $U$ ensures the absence of tachyonic KK gravitational modes. 

Besides the spectrum stability, the potential in Eq. (\ref{potential}) allows a massless KK mode of form 
\begin{eqnarray}
\Psi_0=N_0e^{\frac{1}{2}(3A+B))-\int K(z)dz},
\end{eqnarray}
where $N_0$ is a normalization constant. From the expression in Eq. (\ref{kz}) the torsion modification encoded in the function $K(z)$ acts only for $f_{TT}\neq 0$.


For $p=1$ in Eq. (\ref{coreA}), the conformal coordinate is given by $z=\sinh(\lambda r)/\lambda$ and the superpotential has the form
\begin{eqnarray}\label{43.l}
H&=&\frac{1}{2}\xi
-\left\{\frac{4^nk(n-1)n\ csch^2\left(\frac{2\rho\ arcsinh(z\lambda)}{\lambda} \right) \left(z\lambda^2\xi\right)^n}{z\xi[1+(z\lambda)^2]}\right\}\nonumber\\
&\times&\frac{sinh\left(\frac{2\rho\ arcsinh(z\lambda)}{\lambda} \right)\zeta-4z[1+(z\lambda)^2]^{\frac{3}{2}}\rho^2\ cosh\left(\frac{2\rho\ arcsinh(z\lambda)}{\lambda} \right)}{16z\lambda^2[1+(z\lambda)^2]\rho\ csch\left(\frac{2\rho\ arcsinh(z\lambda)}{\lambda} \right)+ [1+(z\lambda)^2]^{\frac{1}{2}}\varrho},\nonumber\\
\end{eqnarray}
where
\begin{eqnarray}
 \xi&=&-\frac{5z\lambda^2}{1+(z\lambda)^2}+\frac{4\rho\ csch\left(\frac{2\rho\ arcsinh(z\lambda)}{\lambda} \right)}{[1+(z\lambda)^2]^{\frac{1}{2}}},\nonumber\\
 \zeta&=&-2[1+3(z\lambda)^2+2(z\lambda)^4]\rho+z\lambda^2[1+(z\lambda)^2]^{\frac{1}{2}}[3+2(z\lambda)^2]sinh\left(\frac{2\rho\ arcsinh(z\lambda)}{\lambda} \right),\nonumber\\
\varrho&=&nk(4z\lambda^2\xi)^n+z^2\left[nk\lambda^2(4z\lambda^2\xi)^n-20\lambda^4\right].
\end{eqnarray}

\begin{figure}
\begin{center}
\begin{tabular}{ccccccccc}
\includegraphics[height=5cm]{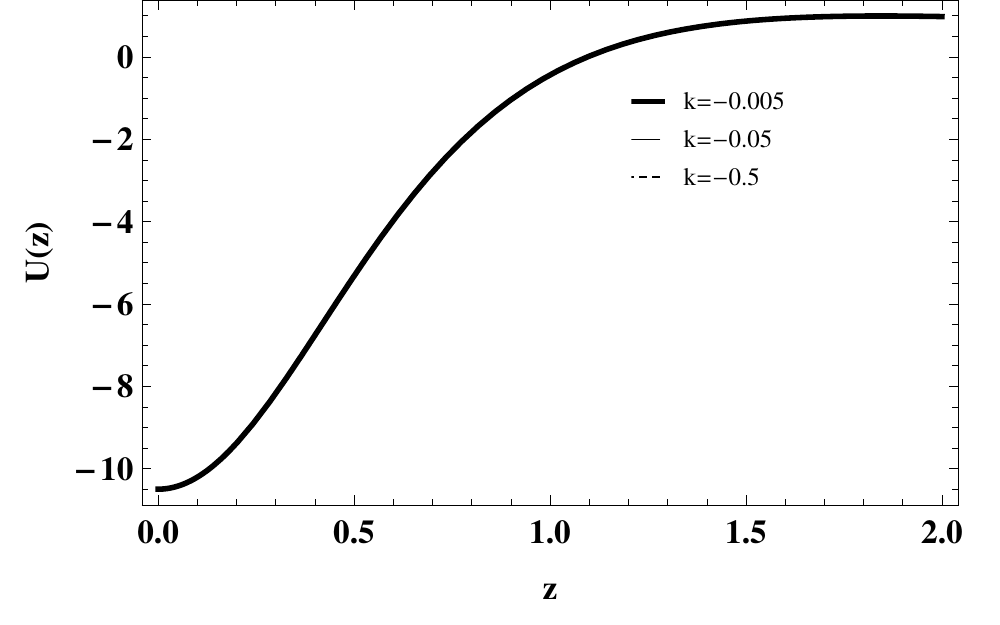} \\
(a)\\
\includegraphics[height=5cm]{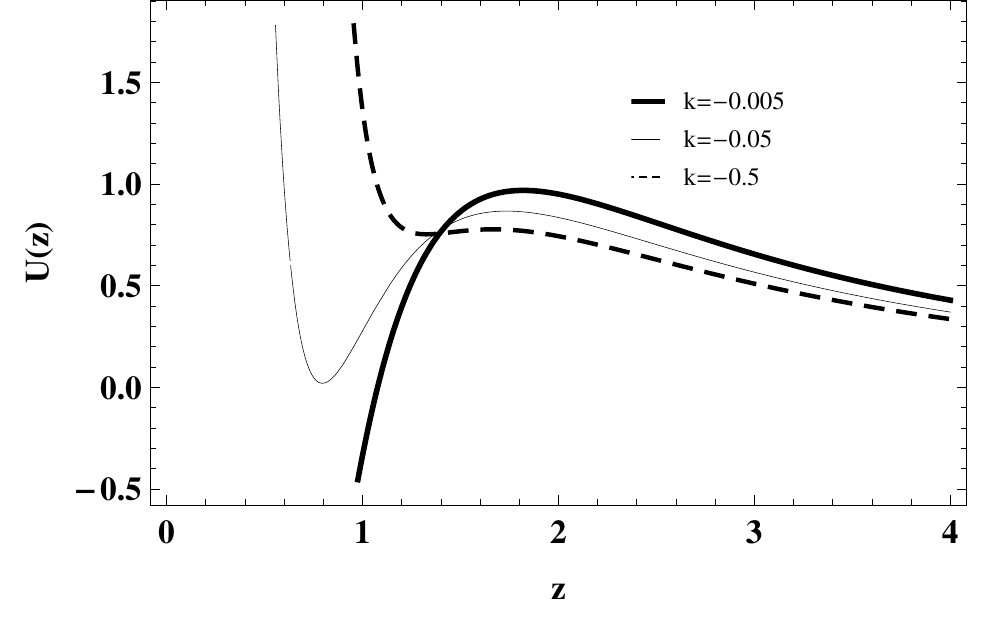} 
\includegraphics[height=5cm]{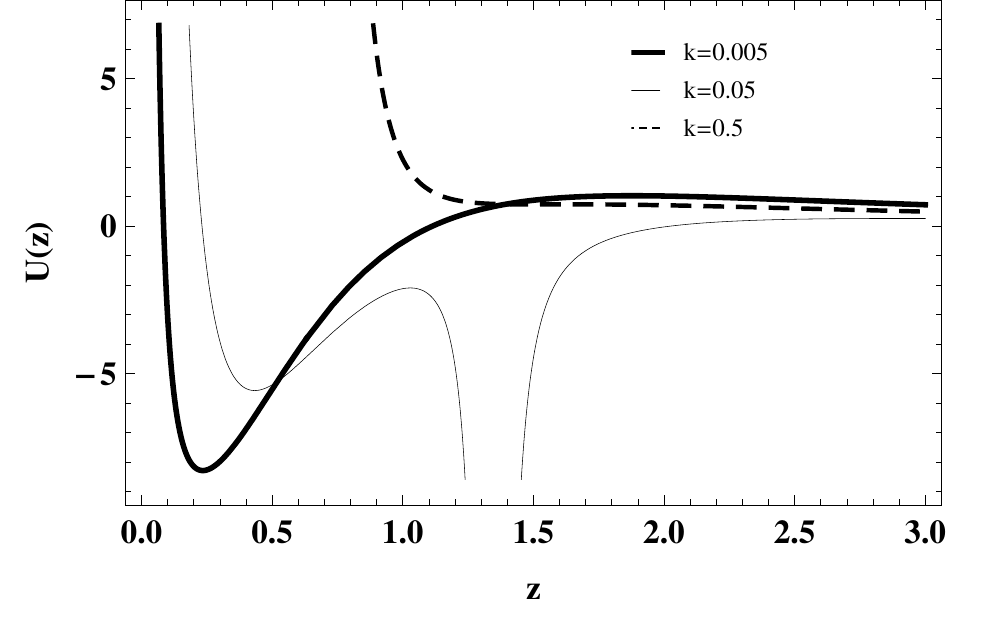}\\
(b) \hspace{8 cm}(c)
\end{tabular}
\end{center}
\caption{ Effective potential for $p=\rho=\lambda=1$. (a) $n=1$. (b)  $n=2$ and $k$ negative. (c) $n=2$ and  $k$ positive. 
\label{figPE1}}
\end{figure}

The expression of the potential is too lengthy to be written here. Instead, we plotted the potential for some values of the torsion parameter and explore some qualitative features. Unlike the GR based string-like branes \cite{conifold,cigar}, for $n=1$ (Fig.\ref{figPE1} ($a$)), the potential has a smooth potential well which is finite at the origin and vanishes asymptotically. Moreover, the effective potential is independent of the torsion parameter $k$. For $n=2$ and $k<0$ (Fig.\ref{figPE1}$(b)$), the potential exhibits a volcano shape with a infinite potential well around the origin, as in the GR based string like models \cite{cigar,regularstring}. However, as $k$ grows in absolute values, the torsion yields to an infinite potential barrier at the origin. For $n=2$ and $k>0$ (Fig.\ref{figPE1}$(c)$), by changing the sign of the $k$, the torsion produces a repulsive potential at the origin and a potential well displayed from the origin.

\begin{figure}
\begin{center}
\begin{tabular}{ccccccccc}
\includegraphics[height=5cm]{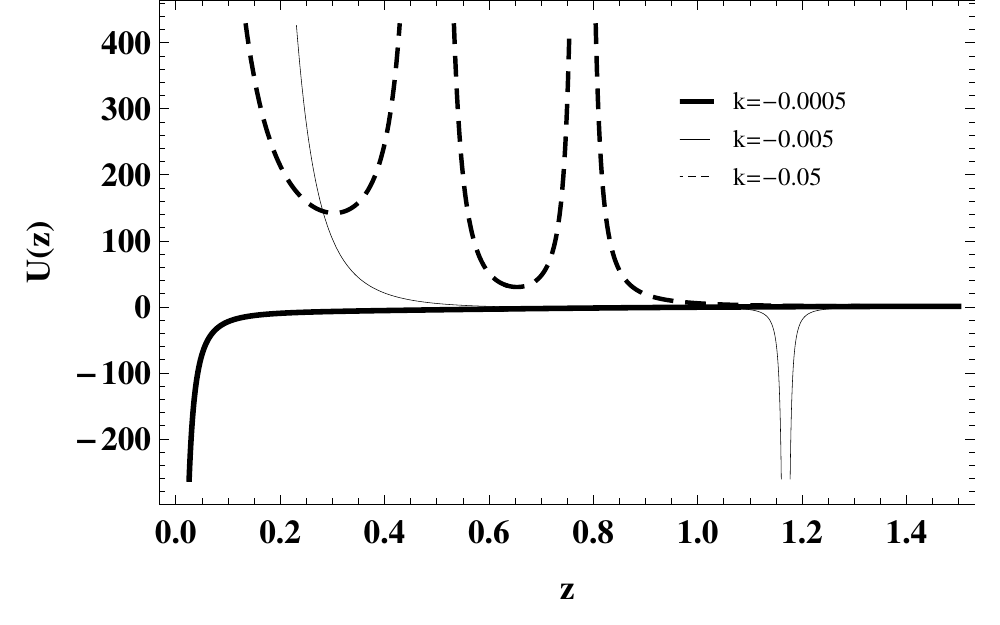} 
\includegraphics[height=5cm]{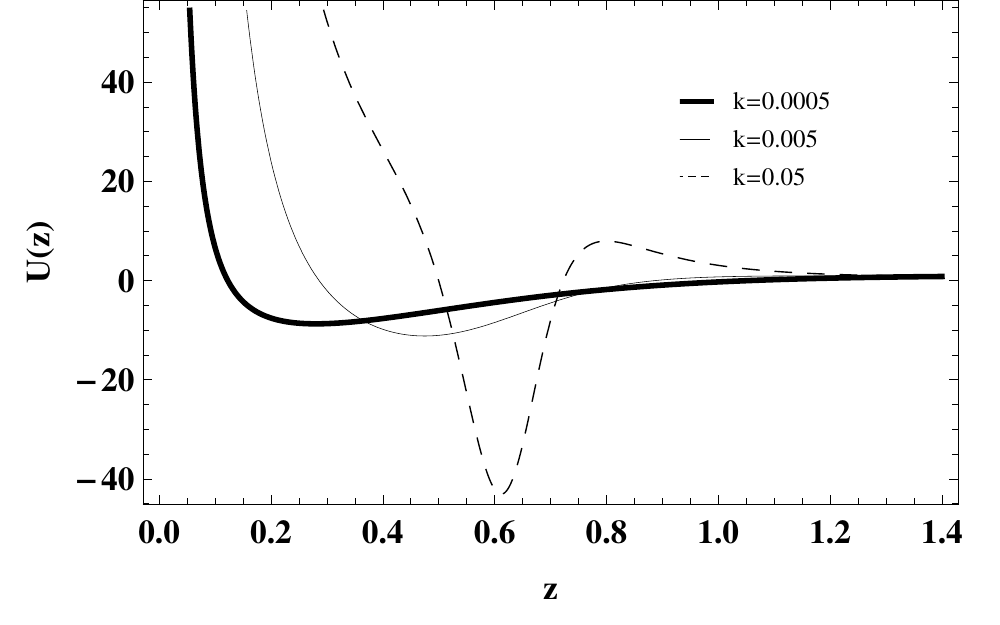}\\
(a) \hspace{8 cm}(b)
\end{tabular}
\end{center}
\caption{Effective potential for $n=3$, $p=\rho=\lambda=1$. (a) $k$ negative 
(b) $k$ positive. 
\label{figPE2}}
\end{figure}

The case $n=3$, plotted in Fig. (\ref{figPE2}), shows additional infinite barriers inside the brane core for $k<0$, as we can see in Fig. (9a). On the other hand, for $k>0$ the torsion provides only a repulsive behaviour at the origin and a potential well shifted from the origin.   

\begin{figure}
\begin{center}
\begin{tabular}{ccc}
\includegraphics[height=5cm]{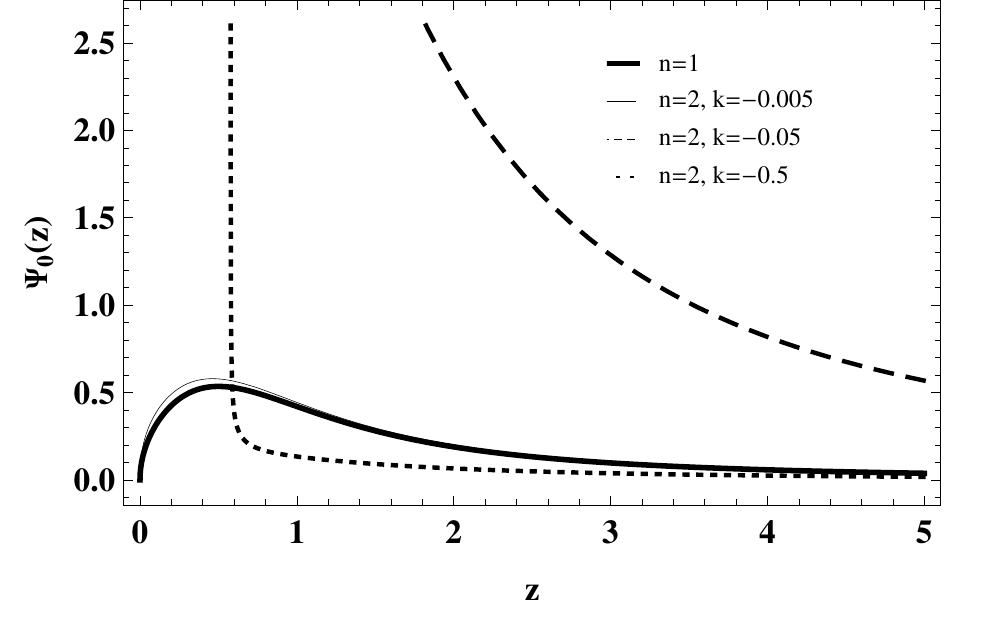} 
\includegraphics[height=5cm]{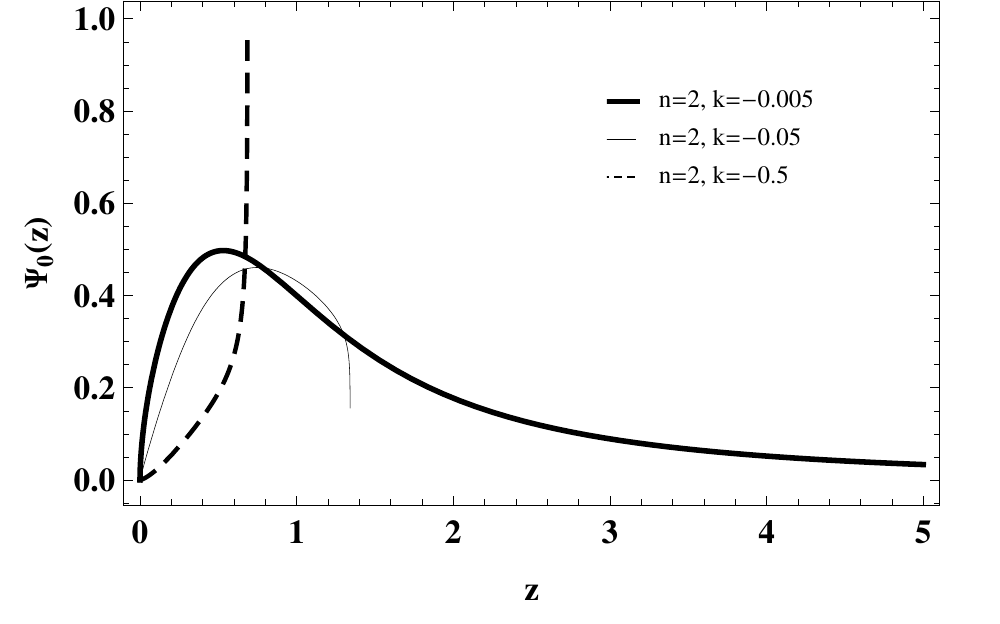}\\
(a) \hspace{8 cm}(b)
\end{tabular}
\end{center}
\caption{Gravitational massless mode for $\rho=\lambda=p=1$. (a) for $k<0$ . 
(b) For $k>0$}. 
\label{masslessmode}
\end{figure}

The profile of the massless mode is depicted in Fig. (\ref{masslessmode}). For $n=1$ the function $K(z)$ vanishes identically and then, the massless mode has the same form as in the GR based string-like braneworld \cite{conifold,cigar,regularstring}. As the torsion increases, the massless mode diverges at the origin. That result agrees with the infinite barrier at the origin and the displayed potential well exhibited by the analogue potential. For $k>0$, the massless mode becomes non-localized as the torsion increases. 


\section{Final remarks and perspectives}
\label{finalremarks}

We studied the torsion effects on a string-like braneworld in the context of the $f(T)$ teleparallel gravity. Likewise the codimension one $f(T)$ braneworld (\cite{Yang2012}), the torsion produces an inner brane structure tending to split the brane.
Furthermore, the $f(T)$ modifications provide a source for the bulk cosmological constant. The exterior region is also modified by the torsion which yields to a AdS-dS transition.

By assuming a single complex scalar field as source, we found usual and deformed global vortex solutions controlled by the torsion parameters. The thick string-like brane undergoes a phase transition evinced by the stress energy components and the torsion invariant. The additional peaks in the energy density at the brane core suggest the formation of a ring-like structure surrounding the initial brane at the origin. Similar behaviour was found in an Abelian local vortex composed by a complex and a vector fields \cite{Giovannini:2001hh} and in a deformed vortex \cite{ringlike}. As the torsion parameters increase, the source violates the dominant energy condition, which reflects on the negative pressure responsible for the brane splitting. A noteworthy extension of the present work is given by the analysis of the solutions for higher winding number and the determination of the corresponding potential. 

The linearized Einstein equation modifies the KK equation only for $f_{TT}\neq 0$. The tower of non-localized massive gravitational fluctuations exhibits a linear gapless spectrum for small masses and a torsion independent behaviour for large masses. For the first massive values, the massive mode has one small bump for $n=1$ and two bumps for $n=3$ inside the brane.
Therefore, the brane splitting process leads to modifications of the massive gravitons inside the thick brane.  

The analysis of the Schr\"{o}dinger-like potential reveals the effects of the torsion on the KK modes. Unlike the GR based string-like braneworld, the torsion removes the divergence of the potential well at the origin for $n=1$, regardless the value of $k$. For $n=2$, the attractive well turns into an infinite barrier as $k$ increases in absolute value. For $k>0$, a finite potential well is formed displayed from the origin. We find an interesting configuration for $n=3$ where multiples wells and infinite barriers are formed shifted from the origin. These barriers prevent the small amplitudes for the massive modes inside the brane. As a result, the torsion tends to localize the gravitational modes in ring-like structures around the origin. Notwithstanding, for $k>0$, the massless mode undergone a transition, becoming non-localized as the torsion increases. These features suggest as an important perspective, the study of the resonant modes and the effects of the KK spectrum on the gravitational potential on the brane.

\section*{Acknowledgments}
\hspace{0.5cm}The authors thank the Conselho Nacional de Desenvolvimento Cient\'{\i}fico e Tecnol\'{o}gico (CNPq), grants n$\textsuperscript{\underline{\scriptsize o}}$ 312356/2017-0 (JEGS) and n$\textsuperscript{\underline{\scriptsize o}}$ 308638/2015-8 (CASA), for financial support.


\end{document}